\documentclass[a4paper,reqno]{amsart}
\usepackage{geometry}
\usepackage{fullpage}
\usepackage[latin1]{inputenc}
\usepackage[T1]{fontenc}
\usepackage{amsfonts}
\usepackage{amssymb}
\usepackage{amsmath}
\usepackage{amsthm}
\usepackage{graphicx}
\usepackage[dvipsnames]{xcolor}
\usepackage{afterpage}
\usepackage[colorlinks=true, linkcolor=magenta, citecolor=cyan, urlcolor=green]{hyperref} %%questo fa i link colorati
\usepackage{bbm}
\usepackage{latexsym}

\numberwithin{equation}{section}

\definecolor{light}{gray}{.9}

\def\be{\begin{equation}}
\def\ee{\end{equation}}
\def\bea{\begin{eqnarray}}
\def\eea{\end{eqnarray}}

\def\nn{\nonumber}
\def\s{\sigma}

\def\a{\alpha}
\def\e{\varepsilon}
\def\epsilon{\e}
\def\d{\delta}
\def\g{\gamma}

\def\b{\beta}

\def\l{\lambda}

\def\t{\tau}

\def\E{\mathbb{E}}

\newcommand{\ie}{\textit{i.e.\ }}

\newcommand{\meanv}[1]{\left\langle#1\right\rangle}

\newcommand{\nocontentsline}[3]{}
\newcommand{\tocless}[2]{\bgroup\let\addcontentsline=\nocontentsline#1{#2}\egroup}

\DeclareMathSymbol{\leqslant}{\mathalpha}{AMSa}{"36} % nicer `smaller or equal'
\DeclareMathSymbol{\geqslant}{\mathalpha}{AMSa}{"3E} % nicer `larger or equal'
\DeclareMathSymbol{\eset}{\mathalpha}{AMSb}{"3F}     % nicer `emptyset'
\renewcommand{\geq}{\;\geqslant\;}                   % redef. of > or =

\newcommand{\tav}{\bar{t}}
\newcommand{\tsqav}{\overline{t^2}}

\title{Phase Diagram of  Restricted Boltzmann Machines and Generalised Hopfield Networks  with Arbitrary Priors}
\date{\today}

\author{Adriano Barra}

\address{Adriano Barra: Dipartimento di Matematica e Fisica Ennio De Giorgi, Universit\`a del Salento, Lecce, Italy}
\email{adriano.barra@unisalento.it}

\author{Giuseppe Genovese}

\address{Giuseppe Genovese: Institut f\"ur Mathematik, Universit\"at Z\"urich,
CH-8057 Z\"urich, Switzerland.}
\email{giuseppe.genovese@math.uzh.ch}

\author{Peter Sollich}

\address{Peter Sollich: Department of Mathematics, King's College London, London WC2R 2LS, UK.}
\email{peter.sollich@kcl.ac.uk}

\author{Daniele Tantari}

\address{Daniele Tantari: Scuola Normale Superiore, Centro Ennio de Giorgi, Piazza dei Cavalieri 3, I-56100 Pisa, Italy.}
\email{daniele.tantari@sns.it}

\begin{document}

\begin{abstract}
Restricted Boltzmann Machines are described by the Gibbs measure of a bipartite spin glass, which in turn corresponds to the one of a generalised Hopfield network. This equivalence allows us to characterise the state of these systems in terms of retrieval capabilities, both at low and high load.
We study the paramagnetic-spin glass and the spin glass-retrieval phase transitions, as the pattern (\ie weight) distribution and spin (\ie unit) priors vary smoothly from Gaussian real variables to Boolean discrete variables. Our analysis shows that the presence of a retrieval phase is robust and not peculiar to the standard Hopfield model with Boolean patterns. The retrieval region is larger when the pattern entries and retrieval units get more peaked and, conversely, when the hidden units acquire a broader prior and therefore have a stronger response to high fields. Moreover, at low load retrieval always exists below some critical temperature, for every pattern distribution ranging from the Boolean to the Gaussian case.
\end{abstract}

%, but broad and high-responding hidden units. Che volevi di?

\maketitle

\section{Introduction}

The genesis of modern AI can be traced quite far back in time. Beyond the pioneering and historical contributions around the beginning of the last century, the most celebrated milestones are the neuron model of McCulloch and Pitts \cite{MC-P}, the Rosenblatt perceptron \cite{Rosenblatt}, and along with the Hebb learning rule \cite{Hebb}. The latter was, in turn, exploited by Hopfield many years later to write his celebrated paper on neural networks from the connectionist perspective \cite{hop}. There has been a growing stream of studies of neural networks ever since, with the subject attracting the interest of various communities, from biological systems to signal processing and information theory \cite{hkp,CKS,engel,DL-book}.  The physics angle on the topic is mainly represented by the statistical mechanics of spin glasses \cite{MPV}. In particular, problems of great biological and technological relevance, such as the capability to learn or retrieve memories, find a simple formulation in a genuine statistical mechanics language \cite{hop,ags1,ags2,hkp,CKS,engel,seung}. 

However, the models used to implement these two crucial features of neural networks -- learning and retrieval -- often start from quite different assumptions. For instance, in modern machine learning approaches such as  {\em deep learning} \cite{deeplearning,DL-book}, network weights are normally taken as real, enabling the use of gradient descent for learning and inference. On the other side, the standard theory of pattern retrieval, as exemplified by the Amit-Gutfreund-Sompolinsky analysis of associative neural networks \cite{ags1,ags2}, assumes Boolean patterns. 
Nevertheless, the two most utilised models for machine learning and retrieval, i.e.\ restricted Boltzmann machines (RBMs) and associative Hopfield networks are known to be equivalent \cite{equivalence,NN?, mezard, huang, gabrie1}. Their relation is easily understood from the point of view of bipartite spin glasses: on the one hand the Gibbs measure of such systems is the same as the one of Restricted Boltzmann Machines, on the other one bipartite spin glasses constitute a class of disordered systems in which the Hopfield model for neural networks can be embedded.

For this reasons in this paper we analyse spin glasses defined on a bipartite network. We study the retrieval in these networks while varying both spin/unit priors and pattern/weight distributions continuously between the Boolean and the real Gaussian limits. We show that the presence of a ferromagnetic region of retrieval is not peculiar to the standard Hopfield model, but it occurs also in the case of continuous units and weights when these take the form of a Gaussian ``softening'' of Boolean variables. Moreover, while retrieval disappears for Gaussian weights at high load, in the low-load limit our generalised Hopfield networks always have a retrieval phase throughout the entire range of pattern distributions ranging from the Boolean to the Gaussian cases. This implies a degree of robustness in the machine-learning set-up, where weights evolve on real axes and one usually works at low load, i.e. with a small number of features, to avoid overfitting \cite{deeplearning, Welling2005}. 

\subsection{Generalised Hopfield Models and Restricted Boltzmann Machines }\label{ss:hopintro1}
The Hopfield model introduced in \cite{hop} is a celebrated paradigm for neural networks in which the neurons are represented by $N$ spins, taking values $\pm 1$. The energy function of the system is defined in terms of $p$ so-called patterns, denoted by $\boldsymbol{\xi}^\mu$, $\mu=1,\dots,p$. It is natural to take the patterns to be $N$-dimensional random vectors with independent and identically distributed components, which makes the Hopfield model a spin glass. Given an instance of the patterns, the Hamiltonian and the Gibbs measure of this system are
\be
H_{N,p}(\s|\xi):=-\sum_{\mu=1}^p N m^2_\mu\label{eq:H-hop} \,,\qquad
G_{N,p}(\s|\xi):=\frac{e^{-\b H_{N,p}(\s|\xi)}}{\E_{\s}e^{-\b H_{N,p}(\s|\xi)}}\,,
\ee
where $\b>0$ is the inverse temperature, $\b =1/T$, $\E_\s$ denotes the statistical expectation with respect to the spin configurations in $\{-1,1\}^N$ and
$$
m_\mu:=\frac1N\sum_{i=1}^N \xi_i^\mu\s_i\,
$$
are the pattern overlaps, or Mattis magnetisations \cite{mattis}. Intuitively, the spin configurations selected by this Hamiltonian have the best possible overlap with the quenched patterns. In particular when the Gibbs average of $m_\mu$ is non-zero for some $\mu$ we say that this pattern is being retrieved. For a short but comprehensive summary of the main known results on this model we refer to section II.B of \cite{mezard}.

A generalisation of the Hopfield model is obtained by replacing $m^2$ in (\ref{eq:H-hop}) with a generic even function $u(m)$:
\be
H_{N,p}(\s|\xi)=-\sum_{\mu=1}^p u(\sqrt{N}m_\mu)\label{eq:H-hop-gen} \,.\qquad
\ee
It is physically interesting, but not necessary, to consider convex $u$ \cite{Rosenblatt,gard,garder,hkp,engel,Talabook}. Any convex, even and smooth $u$ can be expressed as the cumulant generating function of a sub-Gaussian symmetric probability distribution with unit variance \cite{multiGT}. Interpreting the random variables with this distributions as ancillary spins, we obtain a correspondence between generalised Hopfield models and bipartite spin glass models. The latter are defined as follows: consider a bipartite system, with one part containing  $N_1$ spins denoted $\s$ and the other $N_2$ spins written as $\tau$. Also let  $N=N_1+N_2$, $\a=N_2/N$ and define the partition function
\be\label{eq:rbm1}
Z_{N_1,N_2}(\b;\boldsymbol{\xi})=\E_{\s,\tau} \exp\left( \sqrt{\frac \b N} \sum_{i=1}^{N_1}\sum_{\mu=1}^{N_2} \xi^{\mu}_i \s_i\tau_\mu\right).
\ee
Setting $u(x)=\ln\E_{\tau_1}e^{x\tau_1}$, the cumulant generating function of the random variable $\tau_1$, and marginalising over all $\tau$, we clearly obtain the partition function of a generalised Hopfield model with interaction $u$, as claimed. Therefore we can think of the $\xi^\mu_i$ as patterns, each entry being independently drawn from $P_\xi(\xi_i^\mu)$.
On the other hand, (\ref{eq:rbm1}) can be viewed as a Restricted Boltzmann Machine, where a layer of visible units $\boldsymbol{\s}$ interacts with a layer of hidden units $\boldsymbol{\tau}$ through the weights $\boldsymbol{\xi}$. 

The standard Hopfield model is recovered when the $\xi$ and the $\s$ are binary and the $\tau_\mu$ are Gaussian variables, but we study in this paper a much larger class of priors $P_\s(\s_i)$, $P_\tau(\tau_\mu)$ and $P_{\xi}(\xi^\mu_i)$. This corresponds in the generalised Hopfield model to varying the pattern distribution, the spin prior and the form of the interaction $u$.  

Here we investigate the general phase diagram,  especially with regards to the existence of a retrieval phase (focusing on single pattern retrieval) and its interplay with the spin glass phase. Similar models of RBMs with generic priors have recently been studied using belief propagation and related methods in \cite{mezard, gabrie1,gabrie2,huang}.

\begin{figure}
\includegraphics[scale=1.2]{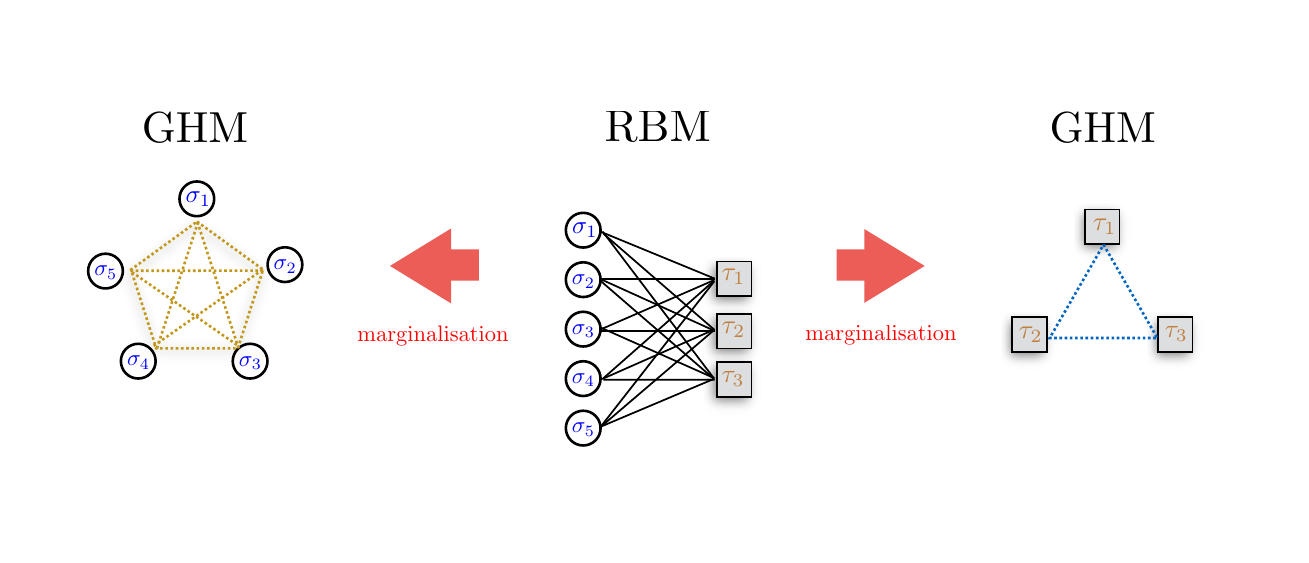}
\caption{Three equivalent architectures of neural networks: in a restricted Boltzmann machine (RBM) (consisting of $N_1=5$ $\s$ variables and $N_2=3$ $\t$ variables in the figure) the role of hidden and visible units can be exchanged and marginalising over the hidden units one gets two dual generalised Holpfield models (GHMs), where the visible layer of the RBM constitutes the network and the hidden layer determines the interaction.}\label{fig0}
\end{figure}

\subsection{Model and RS Equations}\label{ss:intro2}
We shall use random variables which interpolate between Gaussian and binary distributions. Let $\Omega\in[0,1]$, $\e$ $g\sim N(0,1)$ and $\e$ be a symmetric random variable taking values $\pm1 $. We define $\zeta$ as
$$
\zeta(\Omega)=\sqrt \Omega g+\sqrt{1-\Omega}\epsilon\,
$$
and we denote by $\mathcal D(\Omega)$ its probability distribution. Of course $\E[\zeta]=0$ and $\E[\zeta^2]=1$ for all $\Omega$.

Throughout we will draw both the patterns and the spins from $\mathcal D(\Omega)$, \ie
$\xi_i^\mu\sim \mathcal D(\Omega_\xi)$, $\s_i\sim \mathcal D(\Omega_\s)$ and $\t_\mu\sim \mathcal D(\Omega_\t)$ for $\Omega_\xi,\Omega_\s,\Omega_\t\in[0,1]$. It will be useful to define the shorthand $\d=\sqrt{1-\Omega_\xi}$.

To allow for retrieval phases in our analysis, we assume there are some numbers $\ell_1$ and $\ell_2$ of condensed patterns with pattern overlaps or Mattis magnetizations 
\begin{eqnarray}
 m^\mu(\boldsymbol{\s})&=&\frac 1 {N_1} \sum_{i=1}^{N_1}\xi^\mu_i \s_i,  \quad \mu=1,\ldots, \ell_2,\\
 n^i(\boldsymbol{\tau})&=&\frac 1 {N_2}\sum_{\mu=1}^{N_2}\xi^\mu_i \tau_\mu, \quad i=1,\ldots,\ell_1
\end{eqnarray}
of order unity.  We consider, for the sake of simplicity, the possible retrieval of a single pattern, i.e. $\ell_1=\ell_2=1$ or pure state ansatz. The general case of mixed states is a straightforward generalisation \cite{CKS} and can be considered a finer characterisation of the retrieval region we are going to describe.  On the other hand, the possible presence of frozen but disordered states (spin glass region) can be described  by introducing the overlaps 
\be
 q(\boldsymbol{\s}^a,\boldsymbol{\s}^b)=\frac 1 {N_1} \sum_{i=1}^{N_1} \s^a_i\s^b_i,  \quad
r(\boldsymbol{\tau}^a,\boldsymbol{\tau}^b)=\frac 1 {N_2} \sum_{\mu=1}^{N_2} \tau^a_\mu\tau^b_\mu, 
\ee
between two configurations $(\boldsymbol{\s}^a,\boldsymbol{\tau}^a)$ and $(\boldsymbol{\s}^b,\boldsymbol{\tau}^b)$ sampled from the Gibbs measure with the same pattern realisation, and the self-overlaps  $Q(\boldsymbol{\s})$ and $R(\boldsymbol{\tau})$ in the case $a=b$. From a fairly standard replica calculation and the replica symmetry assumption (see Appendix \ref{App-replica} for more details), one gets that in the thermodynamic limit the Gibbs averages of the order parameters converge to the solutions of the following system:
\begin{eqnarray}
m &=& \meanv{\xi \meanv{\s}_{\s|z,\xi}}_{z,\xi} \label{eqs:0}\\
n &=& \meanv{\xi \meanv{\tau}_{\tau|\eta,\xi}}_{\eta,\xi}\label{eqs:1}\\
q         &=& \meanv{ \meanv{\s}^2_{\s|z,\xi}}_{z,\xi} \label{eqs:2}\\
r      &=&   \meanv{ \meanv{\tau}^2_{\tau |\eta,\xi}}_{\eta,\xi}\label{eqs:3} \\
Q         &=& \meanv{ \meanv{\s^2}_{\s|z,\xi}}_{z,\xi} \label{eqs:4} \\
R      &=&   \meanv{ \meanv{\tau^2}_{\tau |\eta,\xi}}_{\eta,\xi} \label{eqs:5}
\end{eqnarray}
Here $z$ and $\eta$ are standard Gaussian random variables, while $\xi$ is sampled \ from $P_\xi$. The distributions of $\s$ and $\tau$ being averaged over are proportional to, respectively,
\begin{eqnarray}
&& P_\s(\s) e^{\b(1-\a)\Omega_\tau  m \xi \s +\sqrt{ \b\a r} \,z\s+{\b \a} (R-r)\s^2/2},\label{eq:effdistr1}\\
&& P_\tau(\tau) e^{ \b\a\Omega_\s n\xi \tau+\sqrt{\b (1-\a) q}\,\eta \tau +{\b(1-\a)}(Q-q) \tau^2 / 2}\label{eq:effdistr2}\,.
\end{eqnarray}
These equations are valid also for more general spin priors $P_\s(\s)$ and $P_\tau(\tau)$, provided one then defines 
$\Omega_{\s}$ (and similarly $\Omega_\tau$) as the high-field response of the spins, in the sense that the average of $\s$ over $P_{\s}(\s)e^{h \s}$  approaches $\Omega_{\s} h$  for large $h$.

We will repeatedly need averages over the distributions (\ref{eq:effdistr1},\ref{eq:effdistr2}). Taking the first as an example, the prior as defined is $P_\s(\s) \propto \sum_\e\exp[-(\s-\e\sqrt{1-\Omega_\s})^2/(2\Omega_\s)]$. Thus the distribution (\ref{eq:effdistr1}) of $\sigma$ has the generic form
\begin{equation}
Z_\s^{-1}\sum_\e e^{-\s^2/(2\gamma_\s)+(\phi_\s\e + h_\s)\s}
\label{eq:effdistr1b}
\end{equation}
where we have set $\phi_\s=\sqrt{1-\Omega_\s}/\Omega_\s$ and
\begin{equation}
\gamma_\s^{-1} = \Omega_\s^{-1} -\b\a(R-r),
\qquad
h_\s = \b(1-\a)\Omega_\tau m\xi + \sqrt{\b\a r} \, z
\label{eq:gamma_and_h_definition}
\end{equation}
Averages over the distribution (\ref{eq:effdistr1b}) then follow from the effective single spin partition function
\be
Z_\s=\int d\s \sum_{\epsilon} e^{-{\s^2}/({2\g_\s})+(\phi_\s\epsilon+h_{\s})\s}\propto \sum_{\epsilon} e^{{\g_\s} (\phi_\s\epsilon+h_{\s})^2/2}\,,
\ee
giving
\begin{eqnarray}
\meanv{\s}_{\s|z,\xi}&=& \partial_{h_\s} \ln Z_\s =\frac{ \sum_{\epsilon} \g_\s(\phi_\s\epsilon+h_{\s}) e^{{\g_\s} (\phi_\s\epsilon+h_{\s})^2/2}}{ \sum_{\epsilon} e^{{\g_\s} (\phi_\s\epsilon+h_{\s})^2/2}} \\
&=& \g_\s h_\s+\g_\s\phi_\s\tanh(\g_\s\phi_\s h_{\s})
\label{eq:sigma_mean}
\end{eqnarray}
The average of $\s^2$ can similarly be found from
\begin{eqnarray}
\meanv{\s^2}_{\s|z,\xi} - 
\meanv{\s}^2_{\s|z,\xi}
&=&\partial^2_{h_\s}\ln Z_\s= 
\partial_{h_\s}\meanv{\s}_{\s|z,\xi} = \g_\s + \g_\s^2\phi_\s^2
[1-\tanh^2(\g_\s\phi_\s h_\s)]
\end{eqnarray}
hence
\begin{eqnarray}
\meanv{\s^2}_{\s|z,\xi}&=&\g_\s +\g_\s^2(h_{\s}^2+\phi_\s^2)+2\g_\s^2\phi_\s h_{\s}\tanh(\g_\s \phi_\s h_\s)\,.
\label{eq:sigma2_mean}
\end{eqnarray}
Analogous results hold for the averages of $\tau$ over the distribution (\ref{eq:effdistr2}).

The RBM and equivalent Hopfield model defined above generalizes a number of existing models that are included as special cases. For $\Omega_\s=0$, $\Omega_\tau=1$ and $\Omega_\xi=0$ we recover the standard Hopfield model, while if $\Omega_\xi=1$ we have the analog Hopfield model studied in \cite{bg,bgg,NN?} (see also \cite{bov-stoc} for the associated Mattis model). For $\Omega_\s=\Omega_\t=0$ we recover the bipartite Sherrington-Kirkpatrick (SK) model studied in \cite{Bip,zigg}. In this case it is known that the thermodynamics is not affected by the pattern distribution \cite{univ}. Throughout this paper we consider only on fully-connected networks: results on the sparse case restricted to the Hopfield model can be found in \cite{multi, agl1, agl2}.

\subsection{Summary and Further Comments}

The aim of this paper is to study the  phase diagram of Restricted Boltzmann Machines with generic priors and pattern/weight distributions as defined above. In general one expects three phases: a high-temperature (or paramagnetic) phase in which the free energy equals its annealed bound and all the order parameters are zero; a glassy phase where all pattern overlaps are still zero but replica symmetry breaking (RSB) is expected and finally a retrieval phase in which the overlap still has a glassy structure, but now one or more pattern overlaps have nonzero mean values. The precise organisation of the thermodynamic states is unknown in the glassy and retrieval regions. In particular, while in the glassy phase it is supposed to be similar to the one of the SK model \cite{MPV,Talabook}, the understanding of the retrieval phase remains severely limited~\cite{CKS,bovbook,Talabook}) and represents a open challenge for theoretical and mathematical physics. 

Throughout the paper the starting point for our analysis will be equations (\ref{eqs:0}--\ref{eqs:5}). We will study them analytically and numerically in the various regimes.

The high-temperature transition is well understood by exact methods for the standard Hopfield model \cite{CKS,bovbook}, for the analog Hopfield model \cite{bg, bgg, NN?} and the bipartite SK model \cite{Bip}. Moving beyond these special cases, in Section \ref{sect:SG-trans} we give a theoretical prediction for the transition of the order parameter $q$ as the distributions of the priors and patterns vary. We will see that the transition is independent of the particular pattern distribution. We find explicit expressions for the transition line for $\Omega_\s=0$ (one layer made of $\pm1$ spins) and (with totally different methods in Appendix \ref{app:sf}) for $\Omega_\s=\Omega_\t=1$.  The remaining intermediate cases are studied by  numerically solving the self-consistency equations (\ref{eqs:0}--\ref{eqs:5}) for the order parameters.

Next we analyse the retrieval region considering the retrieval of one single pattern. A simple argument shows that no retrieval is possible for $\Omega_\s=\Omega_\t=0$: retrieval requires giving up an $O(N)$ amount of entropy in the $\sigma$-system. This is worthwhile only if we can gain an extensive amount of energy. The pattern being retrieved gives a field of $O( \sqrt N)$ acting on a $\t$ spin, so the response of it needs to be also $O(\sqrt{N})$ to get an overall $O(N)$ energy gain. This is impossible for binary $\t$, but {\em is} possible for $\t$ with a Gaussian tail, for which the cumulant generating function grows quadratically at infinity. Hence we always consider $\Omega_\t>0$.

In Section \ref{sect:low-load} we look at the low-load regime in which $\a=0,1$. It turns out that the transitions in $m$ and the replica overlap $q$ occur at the same temperature $T=\Omega_\t$, and both transitions are continuous (as is known to be correct for the standard Hopfield model \cite{CKS, bovbook}). First we concern ourselves with binary spins $\Omega_\s=0$, then to analyse $\Omega_\s>0$ where it turns out that we need to add an appropriate spherical cut-off.

In Section \ref{sect:high-load} we study numerically the retrieval transition at high load, \ie $\a\in(0,1)$ so that the number of patterns is proportional to the system size. First we vary separately $\Omega_\t$ and $\Omega_\xi$ while keeping $\Omega_\s=0$ fixed. At $\Omega_\t=1$ we see absence of retrieval in this analog Hopfield model ($\Omega_\xi=1$), as expected from \cite{bgg}. An analysis at $T=0$ shows, furthermore, that the most efficient retrieval is given by the standard Hopfield model. 

Moving on to $\Omega_\s>0$ we find that the model is well-defined only for high temperature. However, it is interesting that while for $\Omega_\s=\Omega_\t=1$ (Gaussian bipartite model) the divergence of the partition function coincides with the glassy transition, in the intermediate cases there is still a region of retrieval in the phase diagram. Finally when we regularise the model, again with a spherical cut-off, we observe a standard retrieval phase, with a reentrant behaviour of the transition line. The latter would suggest a RSB scenario, as in the standard Hopfield model \cite{reentrant}.

In Appendix \ref{App-replica} we derive equations (\ref{eqs:0}-- \ref{eqs:5}) and in Appendix \ref{app:sf} we briefly analyse the Gaussian bipartite spin glass via Legendre duality, a method introduced for the spherical spin glass in \cite{GT}.

The model we analyse is, for $\Omega_\s$, a neural network with soft spins. Soft spin networks were introduced at an early stage of the development of the field by Hopfield in \cite{hop-grad}, but then were not much studied. From the (bipartite) spin glass perspective, soft spins (spherical or Gaussian) permit analytic methods to be more easily applied, compared to the more commonly studied binary $\pm1$ spins. Indeed there is a substantial number of results in the literature. In \cite{IPC} and \cite{ST} (see also \cite{Talabook}) two similar models of spherical neural networks are introduced, with spherical spins and quadratic(-like) interactions. The authors find the free energy to be RS and no retrieval region. However in \cite{IPC} it is noted that retrieval appears when a quartic term is added to the Hamiltonian. More recently in \cite{baik-lee} a spherical spin glass model was considered with random interaction given by a Wishart random matrix, which is much related with the work in \cite{IPC,ST,Talabook}. The authors find the free energy (which one can argue to be RS by comparison with the Wigner matrix case \cite{BDG,BGGT,GT}) and its fluctuations for all temperatures. No retrieval is observed. Finally a spherical bipartite spin glass is analysed in \cite{ACh} for high temperatures, far from the critical point, and the authors find the free energy in a variational form. Interestingly enough, for this model our analysis yields the same paramagnetic/spin glass transition line as for the bipartite SK model, and no retrieval (see the sections \ref{ss:soft}, \ref{ss:spherical-constr} and Appendix \ref{app:sf}). 

As for the RSB scenario there are only a few results for bipartite models. To the best of our knowledge these are limited to 1RSB for the standard Hopfield model (see \cite{crisanti, CKS}) and to a partial mathematical investigation of the bipartite (in fact multipartite) SK model~\cite{zigg, pan}. Therefore we will restrict ourselves only to RS approximations, when needed.

\section{Transition to the spin glass phase}\label{sect:SG-trans}

At very high temperature ($\b=0$) the distributions (\ref{eq:effdistr1}), (\ref{eq:effdistr2}) have no external effective fields and the thermodynamic state is completely random, with order parameters  $m=n=q=r=0$. Lowering the temperature, a spin glass transition to a frozen but disordered states takes place, creating nonzero overlaps $q$ and $r$ while $m$ and $n$ remain zero. Assuming this transition is continuous, we can linearise equations (\ref{eqs:2},\ref{eqs:3}) for small $q$ and $r$:
\begin{eqnarray}
q&\sim& \b\a r \meanv{\s^2}_{0}^2 + o(r)\,,\\
r&\sim& \b(1-\a) q \meanv{\tau^2}_0^2 + o(q)\,.
\end{eqnarray}
Here $\meanv{\,}_0$ denotes the expectation value w.r.t. (\ref{eq:effdistr1}) and (\ref{eq:effdistr2}) with $q=r=0$ (in particular without the random field). The resulting transition criterion is
\be\label{eq:tc}
1= \b^2 \a (1-\a)  \meanv{\s^2}_0^2 \meanv{\tau^2}_0^2=\b^2 \a(1-\a) Q^2R^2\,,
\ee
where, using $(\ref{eqs:4})$ and ($\ref{eqs:5}$), $Q$ and $R$ are the solutions of
\begin{eqnarray}
Q         &=&  \meanv{\s^2}_0 =  Z_\s^{-1}\E_\s \s^2  e^{{\b\a R}\s^2/2}\,,\\
R      &=&   \meanv{\tau^2}_0 = Z_\tau^{-1}\E_\tau \tau^2  e^{{\b(1-\a) Q}\tau^2/2}\,.
\end{eqnarray}
This result does not depend on the particular pattern distribution $P_\xi(\xi)$ (see also \cite{jsplongo}), but it does clearly involve the spin priors.
With these priors fixed, the transition takes place at an inverse temperature $\b_c(\a)>0$ that is a function of $\alpha$. For $\b<\b_c(\a)$ one finds that the self overlaps are the solutions of
\begin{eqnarray}
Q &=&(1-\b\a\Omega_\s^2R)/(1-\b\a\Omega_\s R)^2\,,\label{eqQ}\\
R &=&(1-\b(1-\a)\Omega_\tau^2Q)/(1-\b(1-\a)\Omega_\tau Q)^2\label{eqR}\,.
\end{eqnarray}
The relation (\ref{eqQ}) can be derived directly from (\ref{eq:sigma2_mean}) with $\g_\s^{-1} = \Omega_\s^{-1}-\b\a R$ and $h_\s=0$, and similarly for (\ref{eqR}). Solving (\ref{eqQ}), (\ref{eqR}) together with $(\ref{eq:tc})$, $T_c(\a)=1/\b_c(\a)$ satisfies
\begin{itemize}
\item[$i)$] $\lim_{\a\to0}T_c(\a)=\Omega_\tau, \quad \lim_{\a\to 1}T_c(\a)=\Omega_\s$\,,
\item[$ii)$] $\lim_{\Omega_{\s}\to 0} T_c(\a) = \frac 1 2 \left(2(1-\a)\Omega_{\tau}+\sqrt{\a(1-\a)} +[\a(1-\a)+4\Omega_{\tau}(1-\Omega_{\tau})(1-\a)\sqrt{\a(1-\a)}]^{1/2}\right)$\,,
\end{itemize}
and of course the symmetric expression for $\Omega_{\t}\to 0$, which is obtained by replacing in $ii)$ $\Omega_{\t}$ by $\Omega_{\s}$ and $\a$ by $1-\a$.

\begin{figure}
\includegraphics[scale=.65]{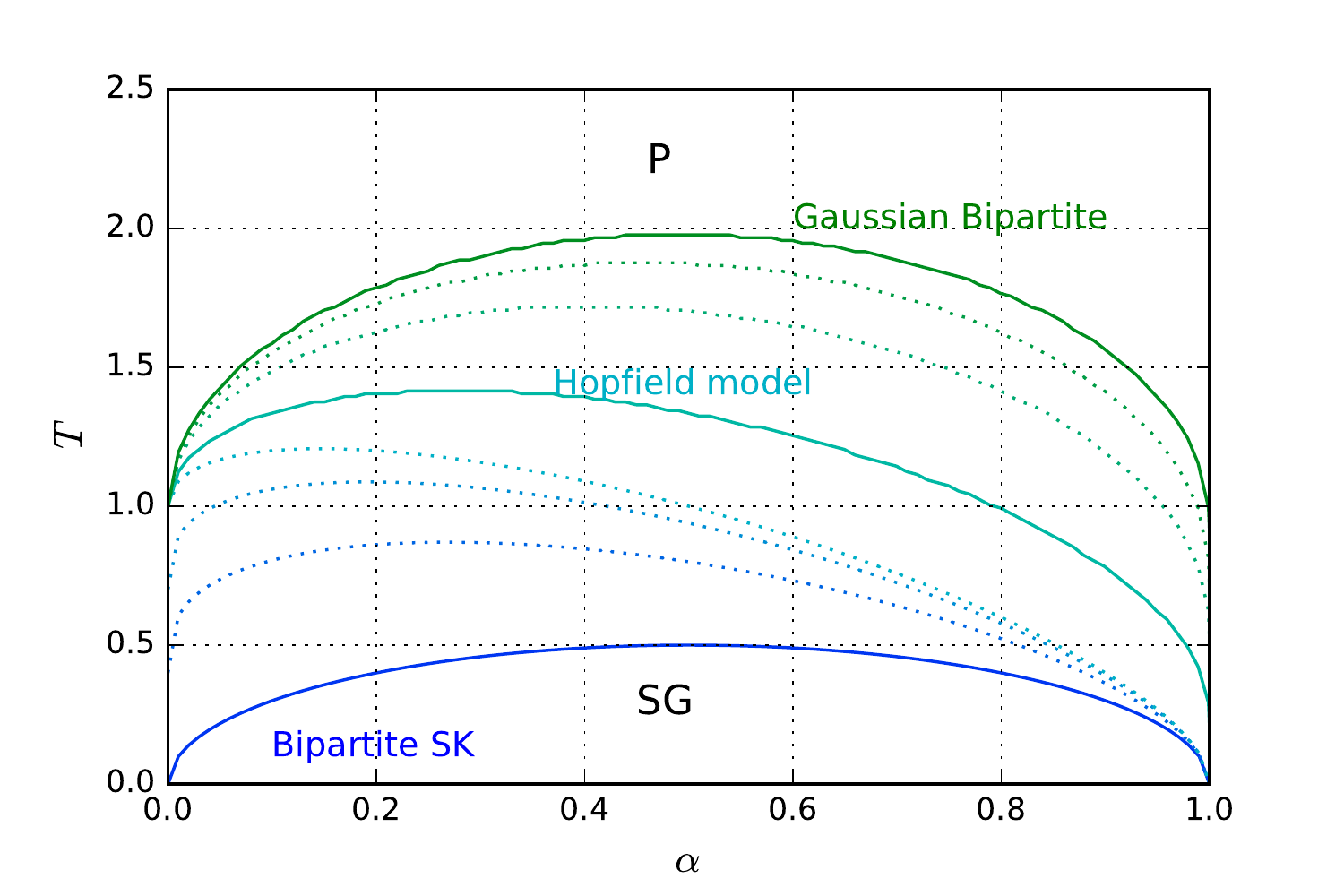}
\caption{Spin glass transition line $T_c(\alpha)$ for different spin priors. The two continuous lines (lower, $\Omega_\s=\Omega_\tau=0$, bipartite SK; upper, $\Omega_\s=\Omega_\tau=1$, bipartite Gaussian) are completely symmetric w.r.t.\ exchange of the two network layers, \ie the transformation $\a\to 1-\a$. In the middle the Hopfield critical line, $\Omega_\s=0$, $\Omega_\tau=1$. % Intermediate values of $\Omega_\tau$ can be read off from the figure simply as $T_c(0)$, and similarly $\Omega_\s=T_c(1)$.
}\label{fig:sgline}
\end{figure}

Relation $ii)$ recovers a number of known special cases. For $\Omega_\s=\Omega_\tau=0$, one gets the critical line of the bipartite SK model $T_c=\sqrt{\a(1-\a)}$ as found in \cite{Bip} (see also \cite{zigg}).  
When $\Omega_\s=0$ and $\Omega_\tau=1$ one has the standard Hopfield model and finds $T_c=1-\a+\sqrt{\a(1-\a)}$ \cite{ags2,NN?}. The case of both Gaussian priors ($\Omega_\s=\Omega_\tau=1$) can be found independently using the Legendre duality between the Gaussian bipartite spin glass model and the spherical Hopfield model studied in \cite{ST,IPC, baik-lee}, see Appendix $\ref{app:sf}$. The general bimodal case can be analysed numerically; the results are shown in Fig.~$\ref{fig:sgline}$.

\section{Transition to retrieval I: low load} \label{sect:low-load}

In the low-load regime the size of one layer is negligible w.r.t.\ the total size of the system, \ie $\a=0,1$. In this case it is possible to obtain equation ($\ref{eqs:0}$) without any RS approximation since the model becomes a generalised ferromagnet. This can be studied only in terms of the pattern overlaps without the need to consider $q$ and $r$ \cite{CKS,bovbook}. Focusing on $\alpha=0$ and linearising ($\ref{eqs:0}$) in $m$ we get
$$
m=\b\Omega_\tau m +O(m^3)\,,
$$
which shows a bifurcation at $T=\Omega_\tau$. As is known special cases, it therefore remains true for generic $\Omega_\s$, $\Omega_\tau$ and $\Omega_\xi$ that the spin glass and low-load retrieval transitions occur at the same temperature.

We next consider the strength of retrieval at temperatures below the transition: the inner average of equation ($\ref{eqs:0}$) is, using (\ref{eq:sigma_mean}) with $\g_\s=\Omega_\s$,
$$\meanv{\s}_{\s|\xi}= \Omega_\s \b \Omega_\tau m \xi +\sqrt{1-\Omega_\s}\tanh(\sqrt{1-\Omega_\s} \b\Omega_\tau m \xi)$$ 
To carry out the remaining average over $\xi$, which by assumption is drawn from the bimodal distribution $\mathcal{D}(\Omega_\xi)$ with peaks at $\pm\d=\pm\sqrt{1-\Omega_\xi}$, we set (see Sec.~\ref{ss:intro2}) $\xi=\d\epsilon + \sqrt{\Omega_\xi} g$. As $\meanv{\s}_{\s|\xi}$ is odd in $\xi$, the two possible values of $\epsilon=\pm 1$ give the same contribution to $\left\langle \xi \langle \sigma \rangle_{\sigma | \xi}\right\rangle$ and we have to average only over $g$. After an integration by parts this gives 
\begin{equation}
m=f_{\b,\boldsymbol{\Omega}}(m)
\end{equation}
with 
\bea
\label{eq:ma0}
f_{\b,\boldsymbol{\Omega}}(m)&=&\b\Omega_\s\Omega_\tau m+\sqrt{1-\Omega_\s} \left\{ \d\,\tav(
\b\sqrt{1-\Omega_\s}\Omega_\tau \d\, m,\sqrt{v})\right.\nn\\
&&{}+\left.
\b\sqrt{1-\Omega_\s}\Omega_\tau \Omega_\xi m
\left[1-\tsqav(
\b\sqrt{1-\Omega_\s}\Omega_\tau \d\, m
,\sqrt{v})\right]  \right\}\nn\,.
\eea
Here have introduced the abbreviations 
\bea
\quad\tav(a,b)=\meanv{\tanh(a+b\,g)}_g\,,&\quad& 
\tsqav(a,b)=\meanv{\tanh^2(a+b\,g)}_g
\label{eq:tbar}
\eea
where the averages are over a zero mean, unit variance Gaussian random variable $g$. We have also defined
\be 
v=\b^2 (1-\Omega_\s)\Omega_\tau^2 \Omega_\xi m^2
\label{eq:vdef1}
\ee 

\begin{figure}
\includegraphics[scale=0.6]{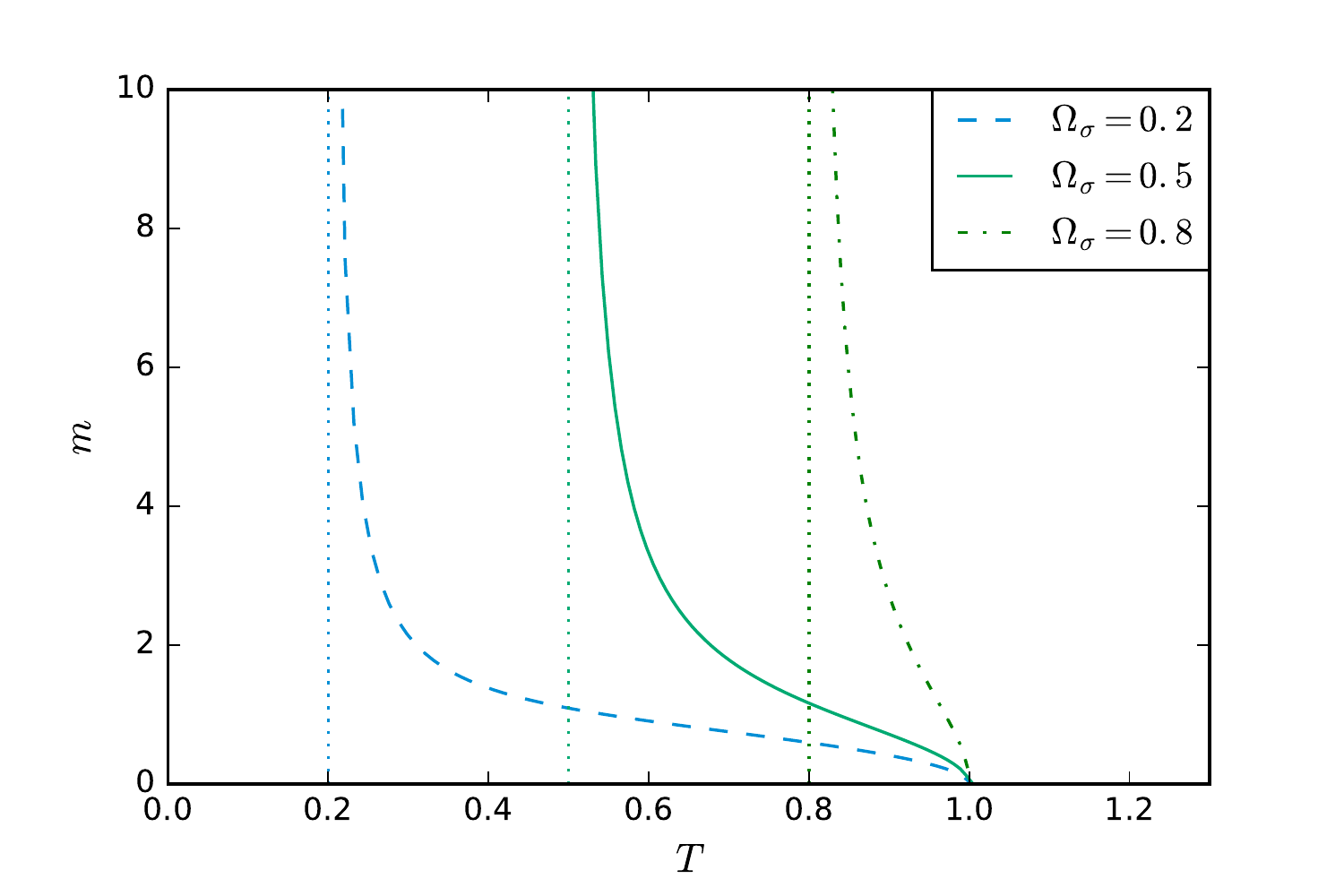}
\caption{Soft bipartite model at low load ($\a=0$): for a generic pattern distribution (here $\d=0.5$) a spontaneous magnetisation appears at $T=\Omega_\tau$, diverging at $T=\Omega_\s\Omega_\tau$.}\label{fig:Gaussa0}
\end{figure}

For binary spins ($\Omega_\s=0$), $|\s|=1$ and so $f_{\b,\boldsymbol{\Omega}}(m)=\left\langle \xi \langle \sigma \rangle_{\sigma | \xi}\right\rangle$ is bounded (between $-\langle |\xi|\rangle$ and $+\langle |\xi|\rangle$). This ensures that a non-trivial solution $m$ of $(\ref{eq:ma0})$ always exists below the retrieval transition. The zero temperature limit of $m$ can be found explicitly: for $\b\to\infty$, $\meanv{\s}_{\s|\xi}\to \operatorname{sgn}(m\xi)$ so $f_{\b,\boldsymbol{\Omega}}(m)\to 
\operatorname{sgn}(m)\meanv{\xi\operatorname{sgn}(\xi)}$ and therefore $m\to \pm \meanv{
|\xi|}$ with
\be\label{eq:ma0t0}
\meanv{|\xi|
} = \sqrt{\frac {2\Omega_\xi} {\pi}} e^{-{\d^2}/({2 \Omega_\xi})}+\d \operatorname{erf} \left(\frac{\d}{\sqrt{2\Omega_\xi} }\right)\,.
\ee
For generic soft spins ($\Omega_\s>0$), on the other hand, $f_{\b,\boldsymbol{\Omega}}(m)$ is no longer bounded but grows as $\b\Omega_\s\Omega_\tau m$ for large $|m|$. The spontaneous magnetisation, which is the solution of $m=f_{\b,\boldsymbol{\Omega}}(m)$, therefore diverges at $T_c=\Omega_\s\Omega_\tau$ as temperature is lowered; 
see Fig.~$\ref{fig:Gaussa0}$. For lower $T$ the model is ill-defined as we are going to see in more detail in the next section, thus we need to regularise the spin distribution in at least one network layer.

\begin{figure}
\includegraphics[scale=0.5]{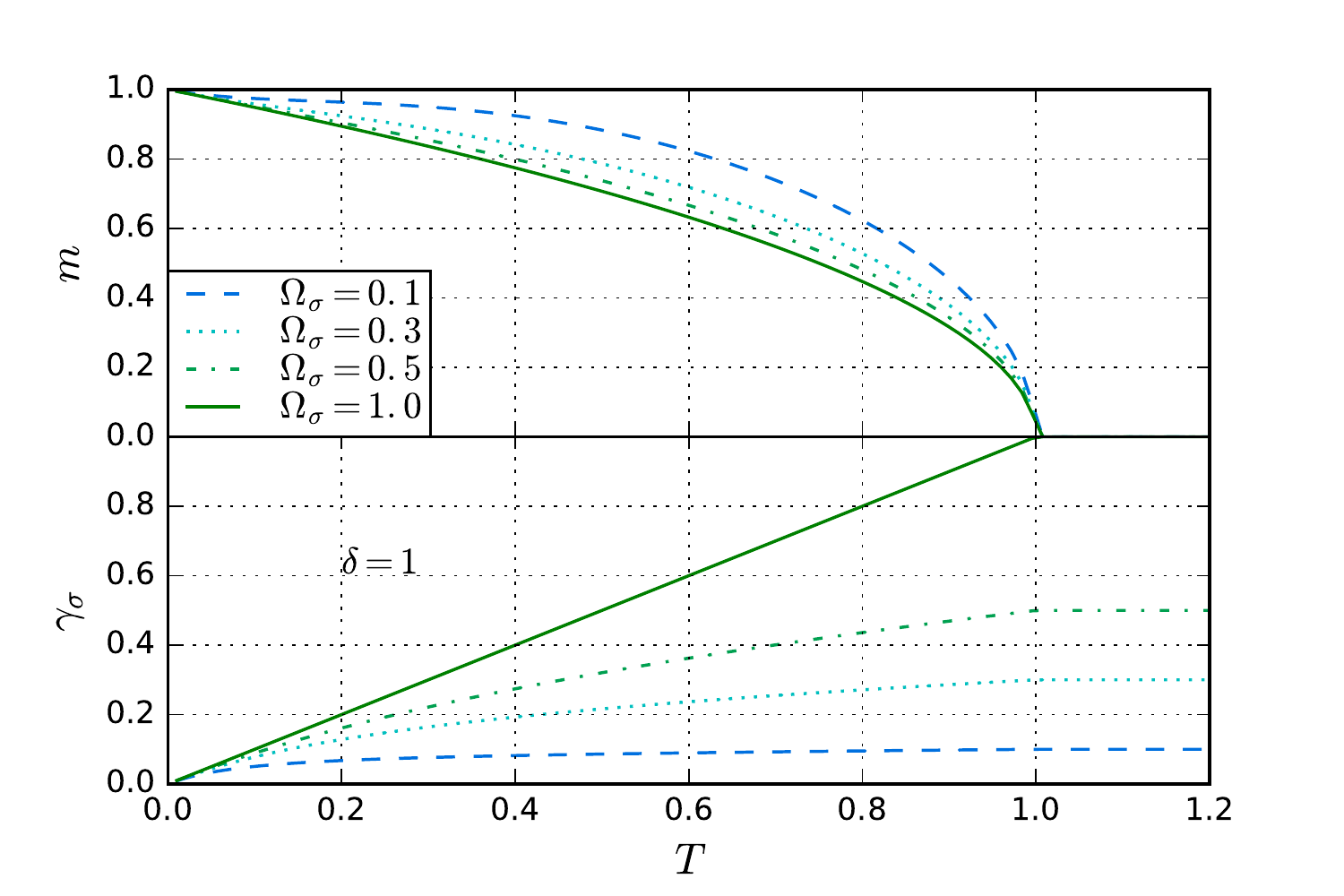}
\includegraphics[scale=0.5]{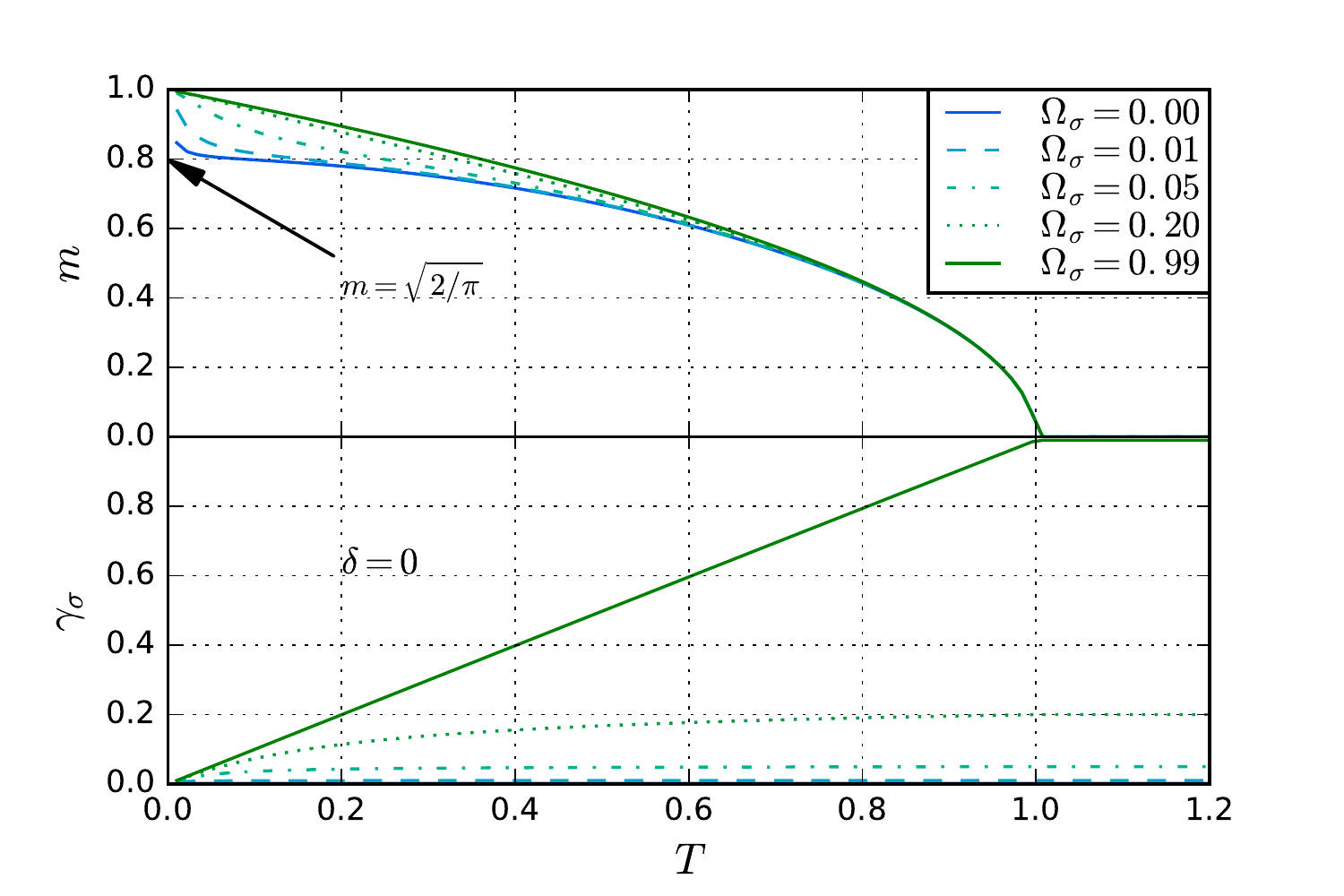}
\caption{Soft model with spherical constraint at low load ($\a=0$). Spontaneous magnetisation still occurs at $T=\Omega_\tau$ increasing until $T=0$. Left panels $\d=1$, right panels $\d=0$. 
As $\Omega_\s\to0$, $m$ approaches the value $ (\ref{eq:ma0t0})$ at low $T$. But at any $\Omega_\s>0$, $m$ eventually peels off from this asymptote to reach $m=1$ for $T\to 0$.
Lower panels  show the behaviour of $\g_\s$: it tends to zero linearly at low temperature, $\g_\s\approx T/\Omega_\tau$, while for $T\geq\Omega_\tau$, $\g_\s=\Omega_\s$. }\label{fig:sfera0}
\end{figure}

To fix the choice of regularisation we note that for a large system, every rotationally invariant weight on the vector of $\s$-spins is equivalent to a rigid constraint at some fixed radius. Without loss of generality we therefore regularize by multiplying the $\sigma$-prior by the spherical constraint $\d(N-\sum_{i=1}^N \s_i^2)$. The resulting prior still depends on $\Omega_\s$; at $\Omega_\s=1$ it is a uniform distribution on the sphere and we obtain the spherical Hopfield model studied in \cite{ST,IPC, baik-lee}. At $\Omega_\s=0$, on the other hand, the regularisation constraint is redundant and we recover the standard Hopfield model. One can now analyse the regularised model using similar replica computations to those above. The only difference is an extra Gaussian factor $e^{-\omega \s^2/2 }$ in the effective $\s$-spin distribution. Here $\omega$ is a Lagrange multiplier that is determined from the spherical constraint $Q=1$. It changes the variance of the two Gaussian peaks fro
 m $\Omega_\s$ to $\g_\s = (\Omega_\s^{-1}+\omega)^{-1}$. 
Accordingly, instead of
$f_{\b,\boldsymbol{\Omega}}$ in $(\ref{eq:ma0})$ one obtains a modified function
\bea
\label{eq:ma0sf}
f_{\b,\boldsymbol{\Omega},\g_\s}(m)&=&\b\g_\s\Omega_\tau m+\g_\s\phi_\s \left\{ \d\,\tav(
\b\g_\s\phi_\s\Omega_\tau \d\, m,\sqrt{v})
+
\b\g_\s\phi_\s\Omega_\tau \Omega_\xi m
\left[1-\tsqav(
\ldots)\right]  \right\}\,.
\eea
where the arguments of $\tsqav$ are the same as for $\tav$. Note that 
the first term of (\ref{eq:ma0}) has become $\b\g_\s\Omega_\tau m$ and all occurrences of $\sqrt{1-\Omega_\s}=\Omega_\s\phi_\s$ have been replaced by $\gamma_\s\phi_\s$. Accordingly, also $v$ now has the more general form
\be
v=\b^2 \g_\s^2 \phi_\s^2
\Omega_\tau^2 \Omega_\xi m^2
\ee
\iffalse
becomes
\begin{eqnarray}\label{eq:ma0sf}
f_{\b,\g_\s}(m)&=&\b\g_\s\Omega_\tau m\\
&&{}+\g_\s\phi_\s \left[ \d\,\tav(
\ldots
) +\b\g_\s\phi_\s\Omega_\tau\Omega_\xi m (1-\bar{t^2} (
\b\gamma_\s\phi_\s\Omega_\tau \d\, m,{v}))  \right]\nonumber\\
m&=&f_{\b,\g_\s}(m)\,,
\end{eqnarray}
where we have simply replaced the Gaussian tail of the priors $\Omega_\s$ with $\g_\s=\Omega_\s/(1+\Omega_\s \omega)$
\fi 
The value of $\omega$ or equivalently $\g_\s$ is defined from the condition $Q=\meanv{\s^2}_{\s,\xi}= 1$, where $Q$ can be worked out using (\ref{eq:sigma2_mean}) as
\be 
Q = \g_\s+\g_\s^2(\b^2\Omega_\tau^2 m^2+\phi_\s^2)
+ 2\b \g_\s^2\phi_\s \Omega_\tau \d\,m\,
\tav(
\ldots
)+ 
2\b^2 \g_\s^3\phi_\s^2 \Omega_\tau^2 \Omega_\xi m^2[1-
\bar{t^2}(\
\ldots)]
\ee
The last two terms are proportional to the last two terms in (\ref{eq:ma0sf}), and hence to $(1-\b\g_\s\Omega_\tau)m$; if one traces back through the derivation this comes from the fact that both results are proportional to $\meanv{h_\s\tanh(\g_\s\phi_\s h_\s)}$. With this simplification one obtains the equivalent expression
\be 
Q =
\label{eq:gamma}
\g_\s +\g_\s^2\phi_\s^2 + \b\g_\s\Omega_\tau m^2(2-\b\g_\s\Omega_\tau) = 1\,.
\ee
\iffalse
\bea
Q &=& \g_\s+\g_\s^2(\b^2\Omega_\tau^2 m^2+\phi_\s^2)
+ 2\b \g_\s^2\phi_\s \Omega_\tau \d\,m\,
\bar{t}(\b \g_\s \phi_\s \Omega_\tau \d\,m,
\b \g_\s \phi_\s \Omega_\tau \d \sqrt{\Omega_\xi}
|m|)
\nonumber
\\&&{}+ 
2\b^2 \g_\s^3\phi_\s^2 \Omega_\tau^2 m^2[1-
\bar{t^2}(\
\ldots,\ldots)]
\eea
\fi
For $\Omega_\s\to 0$ one has $\g_\s\approx \Omega_\s$, which vanishes as $\Omega_\s\to 0$ while $\g_\s\phi_\s = \sqrt{1-\Omega_\s}\g_\s/\Omega_\s \to 1$. For this limiting case of Boolean $\s$-spins the constraint ($\ref{eq:gamma}$) is therefore automatically satisfied as expected.
More generally, while as $m\to\infty$ it behaves as $\b\g_\s\Omega_\tau m$ -- this first term is not the leading contribution because $\g_\s \sim 1/m^2$ for large $m$. The last two terms in $f$ give a nonzero constant asymptote. Therefore $f_{\b,\boldsymbol{\Omega},\g_\s}(m)$ goes as $\b\Omega_\tau (\g_\s+\g_\s^2\phi_\s^2)m$ near $m=0$. From equation ($\ref{eq:gamma}$),
$\g_\s+\g_\s^2\phi_\s^2=1+\mathcal{O}(m^2)$, thus the ferromagnetic transition remains at $T_c=\Omega_\tau$ in the model with the spherical constraint. (One easily checks that $\g_\s+\g_\s^2\phi_\s^2=1$ implies as the physical solution $\g_\s=\Omega_\s$, so that the regularizer $\omega$ increases smoothly from zero at the transition.) For temperatures below $T_c$ one generally has to find $m$ numerically. Results are shown in Fig.~$\ref{fig:sfera0}$. As expected for a regularized model, $m$ remains finite at all $T$. In the low-temperature limit it always reaches its maximum value $m\to 1$. 
One can easily check this from (\ref{eq:ma0sf}) and (\ref{eq:gamma}): the latter implies for $m=1$ that $\b\g_\s\Omega_\tau\to 1$ (see the lower plots in Fig.~$\ref{fig:sfera0}$). Hence the first term on the r.h.s.\ of (\ref{eq:ma0sf}) also approaches unity as it should from $m=f_{\b,\boldsymbol{\Omega},\g_\s}(m)$ while the other terms in (\ref{eq:ma0sf}) vanish in the limit.

\section{Transition to retrieval II: high load}\label{sect:high-load}
Now we study the entire phase diagram of the model, in particular with regards to the presence and stability of a retrieval region. 
We now use the full definition of $\g_\s$ and $h_\s$ from (\ref{eq:gamma_and_h_definition}), along with the analogous definition for $\g_\tau$:
\be\label{eq:gamma_s,t}
\g_\s^{-1}=\Omega_\s^{-1} -\b\a(R-r)\,,\quad \g_\t^{-1}=\Omega_\t^{-1}-\b(1-\a)(Q-q)\,,
\quad
h_\s=\b(1-\a)\Omega_\tau m\xi+\sqrt{\b\a r}\,z\,.
\ee
\iffalse
Then we can unroll the equations ($\ref{eqs:0}-\ref{eqs:5}$) by defining the effective single spin partition function
\be
Z_\l=\int d\s \sum_{\epsilon} e^{-\frac{\s^2}{2\g_\s}+(\phi_\s\epsilon+H_{\s} +\l)\s}\propto \sum_{\epsilon} e^{\frac{\g_\s}{2} (\phi_\s\epsilon+H_{\s} +\l)^2}\,.
\ee
Thus we get
\begin{eqnarray}
\meanv{\s}_{\s|z,\xi}&=& \partial_\l \ln Z_\l|_{\l=0}=\frac{ \sum_{\epsilon} \g_\s(\phi_\s\epsilon+H_{\s} +\l) e^{\frac{\g_\s}{2} (\phi_\s\epsilon+H_{\s} +\l)^2}}{ \sum_{\epsilon} e^{\frac{\g_\s}{2} (\phi_\s\epsilon+H_{\s} +\l)^2}}|_{\l=0} \nonumber \\
&=& \g_\s H_\s+\g_\s\phi_\s\tanh(\g_\s\phi_\s H_{\s})\nonumber\\
\operatorname{Var}_{\s|z,\xi}(\s)&=&\partial^2_{\l^2}\ln Z_\l= -\frac{(\partial_\l Z_\l)^2}{Z_\l^2}+\frac{ \sum_{\epsilon} \g_\s(1+\g_\s(\phi_\s\epsilon+H_{\s} +\l)^2) e^{\frac{\g_\s}{2} (\phi_\s\epsilon+H_\s +\l)^2}}{ \sum_{\epsilon} e^{\frac{\g_\s}{2} (\phi_\s\epsilon+H_{\s} +\l)^2}}|_{\l=0} \nonumber \\
&=&  -\meanv{\s}_{\s|z,\xi}^2+\g_\s +\g_\s^2(H_{\s}^2+\phi_\s^2)+2\g_\s^2\phi_\s H_{\s}\tanh(\g_\s \phi_\s H_\s), \nonumber\\
\meanv{\s^2}_{\s|z,\xi}&=&\g_\s +\g_\s^2(H_{\s}^2+\phi_\s^2)+2\g_\s^2\phi_\s H_{\s}\tanh(\g_\s \phi_\s H_\s)\,\nn.
\end{eqnarray}
\fi
Furthermore we abbreviate the variance of the Gaussian part of $\g_\s\phi_\s h_\s$ as 
\be
v=\b^2(1-\a)^2
\g_\s^2 \phi_\s^2
\Omega_\tau^2\Omega_\xi m^2 +\b\a\g_\s^2 \phi_\s^2 r
\ee
where compared to (\ref{eq:vdef1}) we again have the replacement of $\sqrt{1-\Omega_\s}$ by $\g_\s\phi_\s$, and otherwise the incorporation of the $\alpha$-dependence and the new term proportional to $r$.
Then, taking the averages w.r.t. $\xi$ and $z$ we have, using (\ref{eq:sigma_mean},\ref{eq:sigma2_mean}) and integrating by parts where appropriate, 
\begin{eqnarray}
m&=&\meanv{\xi\meanv{\s}_{\s|z,\xi}}_{z,\xi}\nonumber\\
&=&
\b(1-\a)\g_\s\Omega_\tau m 
+\g_\s\phi_\s \left[ 
\d\,\tav
(\b(1-\a)\gamma_\s \phi_\s\Omega_\tau \d\,
m,\sqrt{v})
+\b(1-\a)\g_\s\phi_\s
\Omega_\tau \Omega_\xi m \left(1-\tsqav(
\ldots
)\right)  \right]\nonumber\\
q&=&\meanv{\meanv{\s}^2_{\s|z,\xi}}_{z,\xi}= \meanv{(\g_\s h_\s+\g_\s\phi_\s\tanh(\g_\s\phi_\s h_\s))^2}_{z,\xi}\nonumber\\
&=& \b^2(1-\a)^2\g_\s^2\Omega_\tau^2 m^2+
\b\a \g_\s^2 r + \g_\s^2\phi_\s^2 \tsqav(\ldots)
+ 2\b(1-\a)\g_\s^2 \phi_\s\Omega_\tau \d\,m
\,\tav(\ldots)
+ 2\g_\s v
\left[1-\tsqav(\ldots)\right]
\nonumber\\
&=& \b(1-\a)\g_\s\Omega_\tau (2-\b(1-\a)\Omega_\tau\g_\s)m^2 + \b\a \g_\s^2(1+2\g_\s\phi_\s^2)r 
+\g_\s^2\phi_\s^2(1-2\b\a \g_\s r)\tsqav(
\ldots)\nonumber\\
Q&=& \meanv{\meanv{\s^2}_{\s|z,\xi}}_{z,\xi}= q + \g_\s +\g_\s^2\phi_\s^2 \left[1-\tsqav(
\ldots)\right]
\nonumber\,.
\end{eqnarray}
where all $\tanh$-averages $\tav$ and $\tsqav$ are evaluated for the same parameters, as given in the equation for $m$. In the final expression for $q$ we have eliminated the $\tav$ term using the expression for $m$.
Repeating the same argument for the effective distribution of the $\tau$ spins, we get the equations for the other order parameters simply by exchanging labels appropriately and replacing $\a$ with $1-\a$, bearing in mind also that the corresponding magnetization parameter is $n=0$. This gives the following additional equations: 
\begin{eqnarray}\label{eq:gen}
r&=&\b(1-\a) \g_\tau^2(1+2\g_\tau\phi_\tau^2)q +\g_\tau^2\phi_\tau^2(1-2\b(1-\a) \g_\tau q)\bar{t^2}(0,\g_\tau\phi_\tau\sqrt{\b(1-\a) q})\label{eq:r}\\
R&=& r+ \g_\tau +\g_\tau^2\phi_\tau^2 
\left[1-
\tsqav(0,\g_\tau\phi_\tau\sqrt{\b(1-\a) q})\right].
\end{eqnarray}

%%%%%%%%%%%%%%%%%%%%%%%%%%%%%%%%%%%%%%%%%%%%%%%

\subsection{One Boolean layer}

In the case where the $\s$-spins are Boolean, $\Omega_\s=0$, the saddle point equations (\ref{eq:gen}) simplify considerably.
From (\ref{eq:gamma_s,t}), one has as before $\g_\s \approx \Omega_\s\to 0$ and $\g_\s\phi_\s \to 1$. This leads to
\begin{eqnarray}\label{eq:sct}
m &=& \d \,\tav(\b(1-\a)\Omega_\tau \d\, m,
\sqrt{v})
+\b(1-\a)\Omega_\tau \Omega_\xi m\left[1-\tsqav(\ldots)\right]
\\
\label{eq:sctq}
q &=& \tsqav(\ldots
)
\end{eqnarray}
where after the inserting the expression (\ref{eq:r}) for $r$ the Gaussian field variance can be written as $v=\b^2(1-\a)^2 V$ with
\be\label{eq:sct2}
V=\Omega_\tau^2 \Omega_\xi m^2 + \frac{\a}{1-\a}\gamma_\tau^2 (1+2\gamma_\tau\phi_\tau^2)q
+\frac{\a}{1-\a}\gamma_\tau^2\phi_\tau^2
[(\b(1-\a))^{-1}-2\gamma_\tau q]\tsqav(0,\gamma_\tau\phi_\tau\sqrt{\b(1-\a) q})\,.
\ee

\begin{figure}
\includegraphics[scale=0.5]{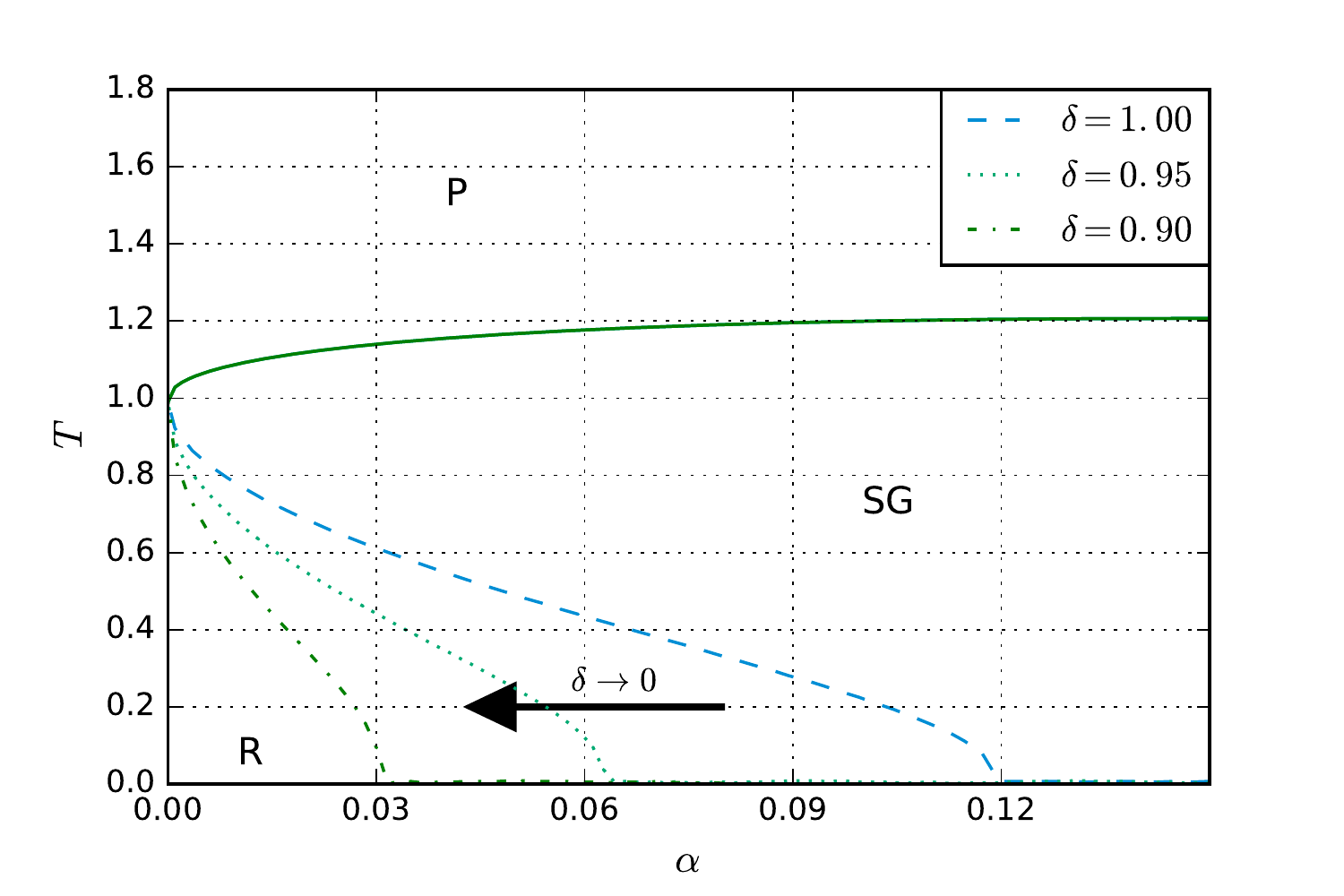}
\includegraphics[scale=0.5]{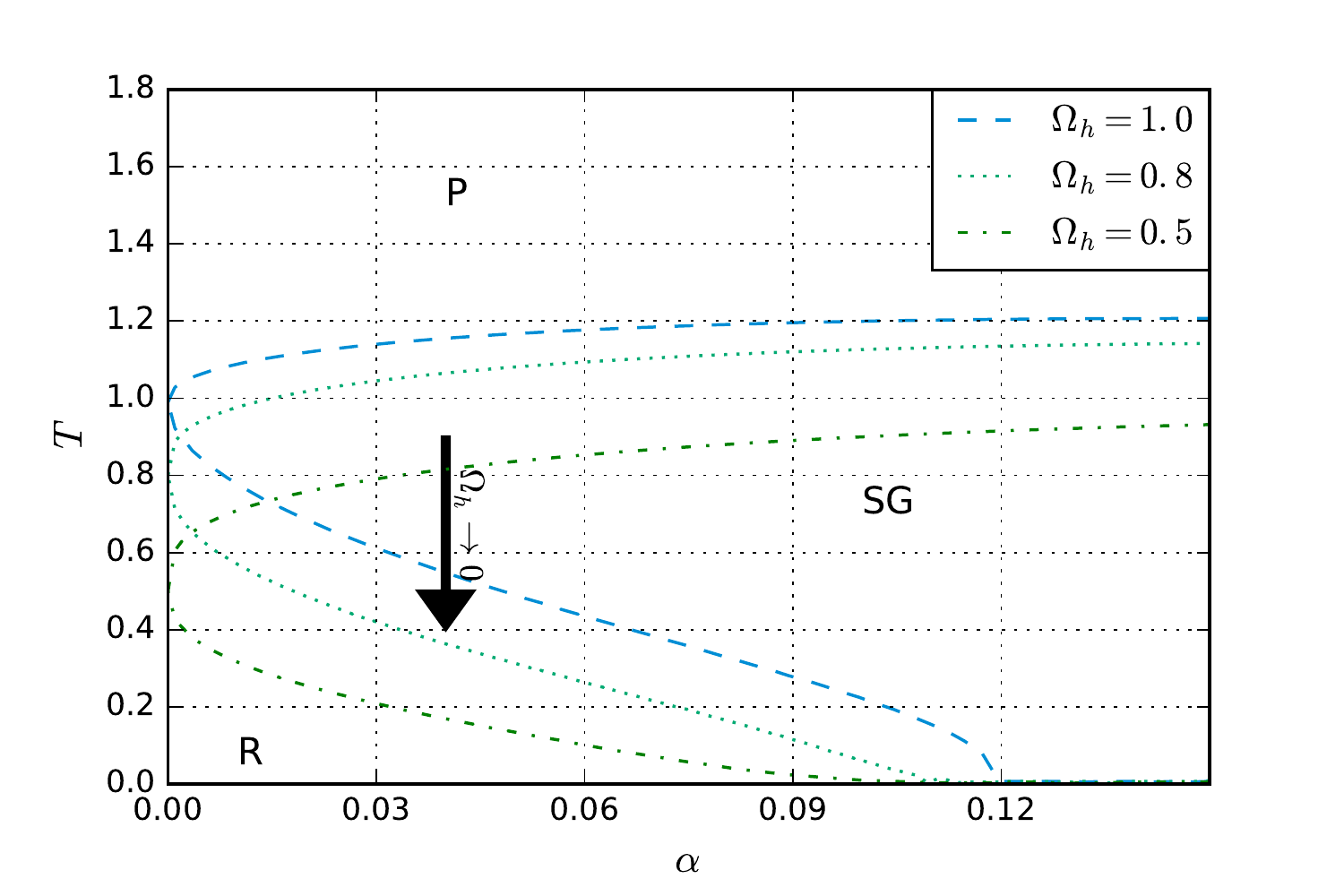}
\caption{Phase diagrams with one Boolean layer ($\Omega_\s=0$). Left panel:  $(T,\a)$ phase diagram for $\Omega_\tau=1$ and different values of $\d$. The retrieval transition line moves towards the $T$-axis as $\d$ decreases while the critical temperature at $\a=0$ remains fixed. Right panel: phase diagram for $\d=1$ and different values of $\Omega_\tau$. Both transition lines move towards the $\a$-axis as $\Omega_\tau$ decreases while now the critical load at $T=0$ is fixed.}\label{fig:pd1}
\end{figure}

Solutions of (\ref{eq:sct}) are shown in Fig.(\ref{fig:pd1}). Starting from the standard Hopfield phase diagram ($\Omega_\xi=0$ and $\Omega_\tau=1$) the retrieval region gradually disappears with decreasing $\Omega_\xi$ or increasing $\Omega_\tau$. In the first case it shifts towards the $T$-axis, as the critical temperature for $\a=0$ is independent of $\Omega_\xi$. In the second case, both the retrieval and spin glass transition lines shift towards the $\a$-axis, as the critical $\a$ at $T=0$ is independent of $\Omega_\tau$ as we will see shortly. 

\subsection{Zero temperature limit}
\begin{figure}
\includegraphics[scale=.5]{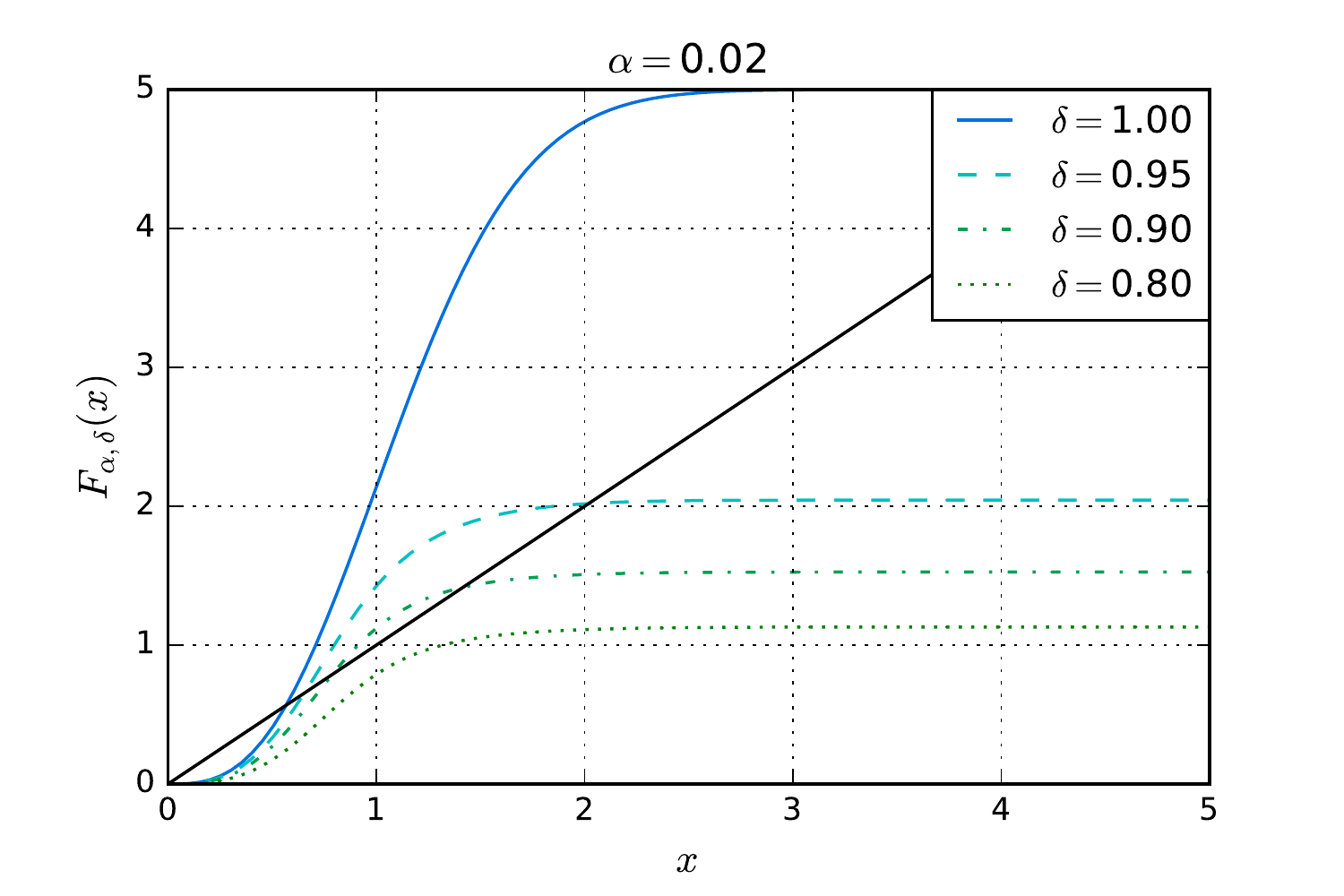}
\includegraphics[scale=.5]{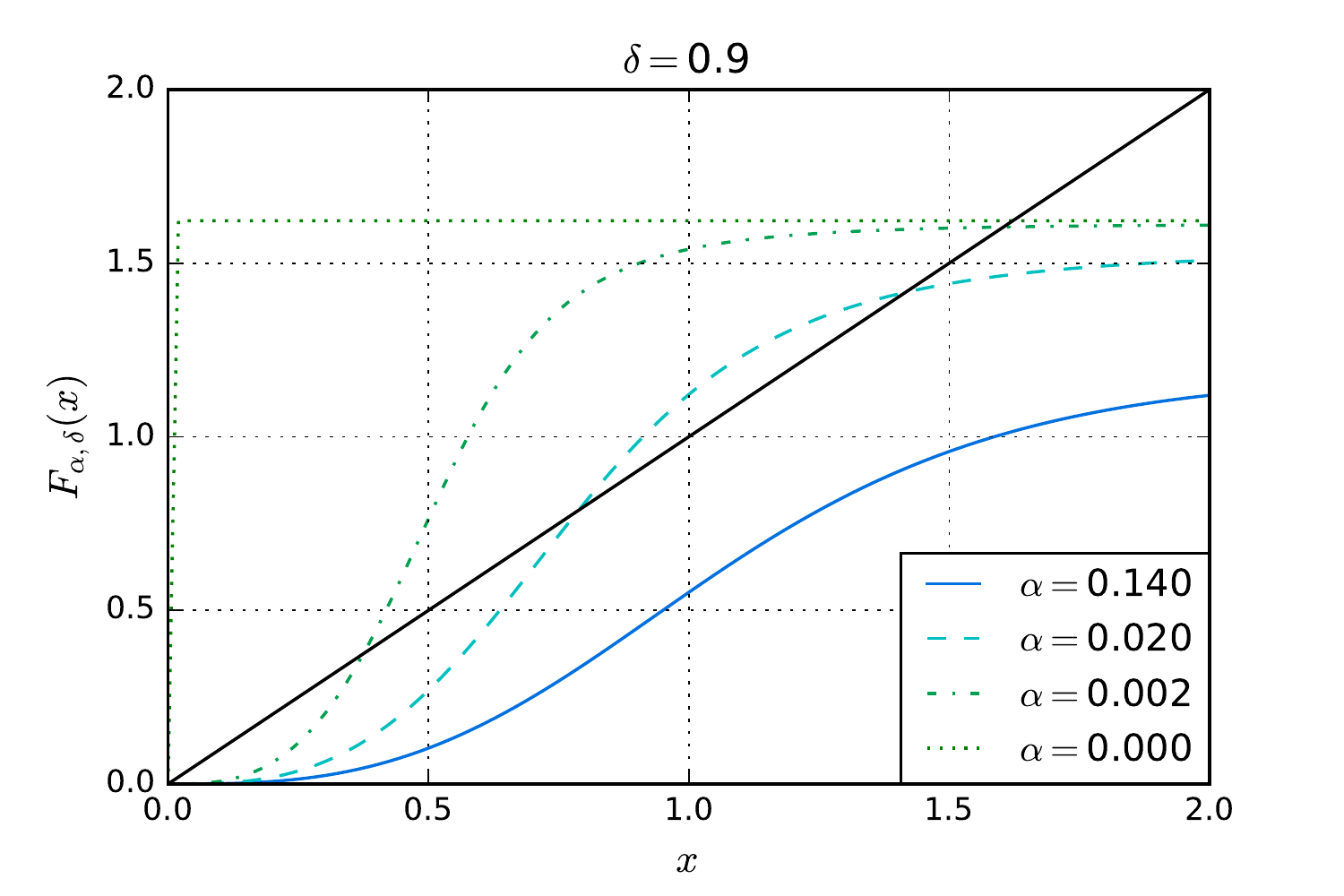}
\caption{Plot of $F_{\d,\a}(x)$. It tends uniformly to zero as $\d \to 0$ at fixed $\a$ (left panel), while it approaches $1/\sqrt{2(1-\d^2)}$ as $\a\to 0$ at fixed $\d$ (right panel). }\label{fig1}
\end{figure}

Useful insight into the $\Omega_\s=0$ case can be obtained by further specializing to the limit $T\to0$ (i.e. as $\b\to\infty$). In this scenario
\be
\bar{t}(\b a,\b b) \to \meanv{\operatorname{sgn}(a+b\eta)}=\operatorname{erf}(a/ \sqrt{2}b)
\ee
and, putting $w=\b(a+b g)$,
\begin{eqnarray}
\b[1-\bar{t^2}(\b a, \b b)] &=& \b\int\frac{dw}{\sqrt{2\pi}\b b} \exp[-(w-\b a)^2/(2\b^2b^2)][1-\tanh^2(w)]\nonumber \\
&\to&  \int\frac{dw}{\sqrt{2\pi} b} \exp[- a^2/(2b^2)][1-\tanh^2(w)]\nonumber \\
&=&\frac{\sqrt{2}}{\sqrt{\pi} b} \exp[-a^2/(2b^2)]\,.
\end{eqnarray}
If we set $v=\b^2(1-\a)^2 V$ as before and then apply the above large-$\b$ identities in the equation ($\ref{eq:sct}$) for $m$ we get 
\be
m=\d\, \operatorname{erf} (\Omega_\tau \d\,m /\sqrt{2 V}) + \Omega_\tau \Omega_\xi m \frac{\sqrt{2}}{\sqrt{\pi V}} \exp(-\Omega_\tau^2 \d^2\, m^2/2V)\,.
\label{eq:m_low_T}
\ee
The equation ($\ref{eq:sctq}$) for $q$ has a limit in terms of $C=\b(1-\a)(1-q)$:
\be
C=\frac{\sqrt{2}}{\sqrt{\pi v}}  \exp(-\Omega_\tau^2 \d^2 m^2/2V)\,.
\ee
Finally for $V$ in ($\ref{eq:sct2}$) the zero temperature limit is simple as $\tsqav(0,\b b) \to 1$ and $q\to 1$, giving
\be
V=\Omega_\tau^2 \Omega_\xi m^2 +\frac{\a}{1-\a}\g_\tau^2= \Omega_\tau^2 \Omega_\xi m^2 +\frac{\a}{1-\a}(\Omega_\tau^{-1} -C)^{-2}\,.
\ee
One can reduce these three equations to a single one for $x= \Omega_\tau m /\sqrt{2V}$, which reads
\be
 x = F_{\d,\a}(x), \qquad
 F_{\d,\a}(x)=\frac{\d \operatorname{erf} (\d x) - \frac{2}{\sqrt{\pi}}x\d^2 e^{-\d^2x^2}}{[2\a +2(1-\d^2)
(\d \operatorname{erf} (\d x) - \frac{2}{\sqrt{\pi}}x\d^2 e^{-\d^2x^2})^2]^{1/2}},
\label{eq:x}
\ee
We leave the derivation of this result to the end of this section.
One sees that $F_{\d,\a}(x)$ is strictly increasing, starting from zero and approaching $\d/\sqrt{2\a +2(1-\d^2)\d^2}$ for large $x$ (Fig.~{\ref{fig1}}). Note also that $\Omega_\t$ has no effect on the value of $x$, and only affects the coefficient in the linear relation between $x$ and $m$.

\begin{figure}
\includegraphics[scale=.5]{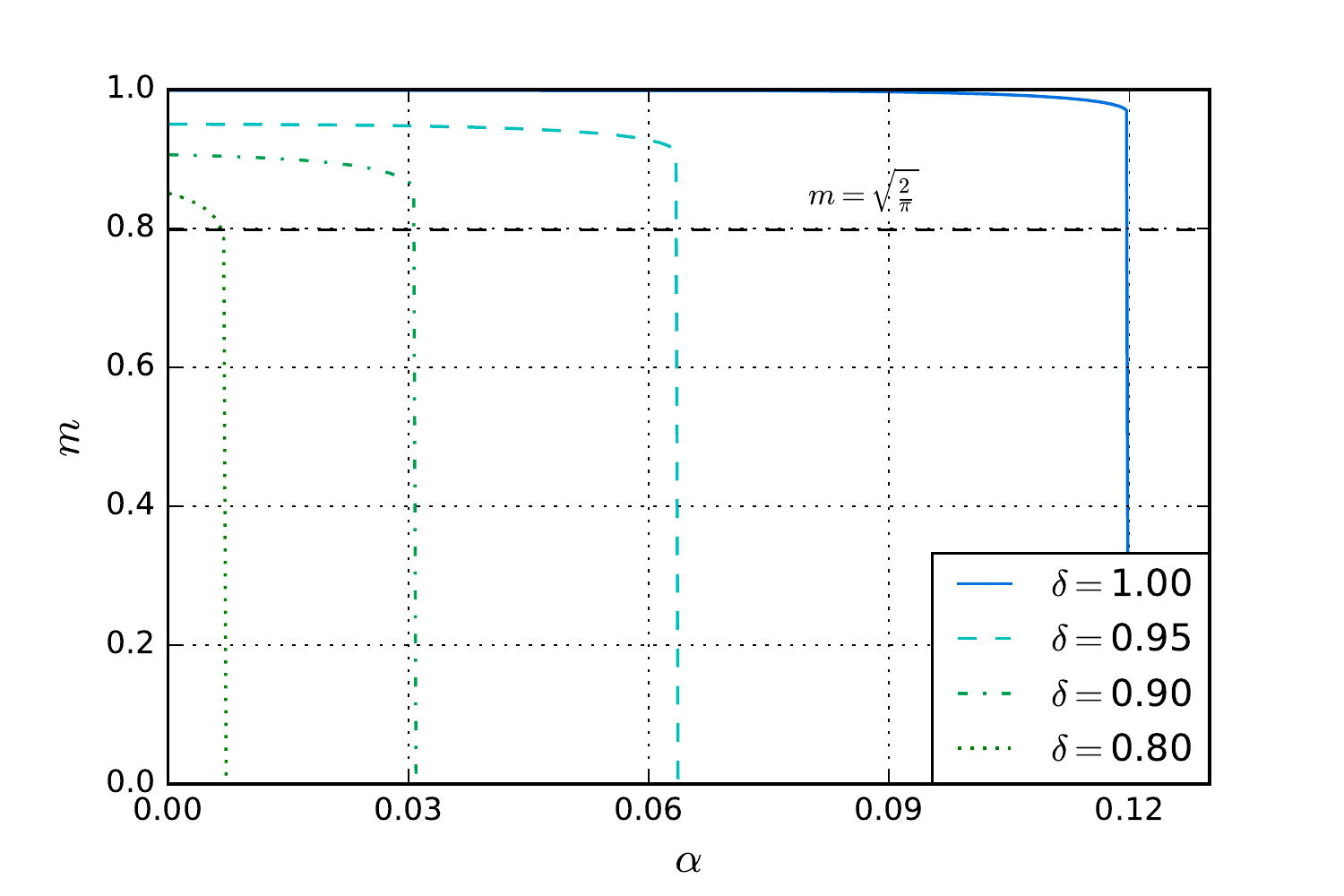}
\includegraphics[scale=.5]{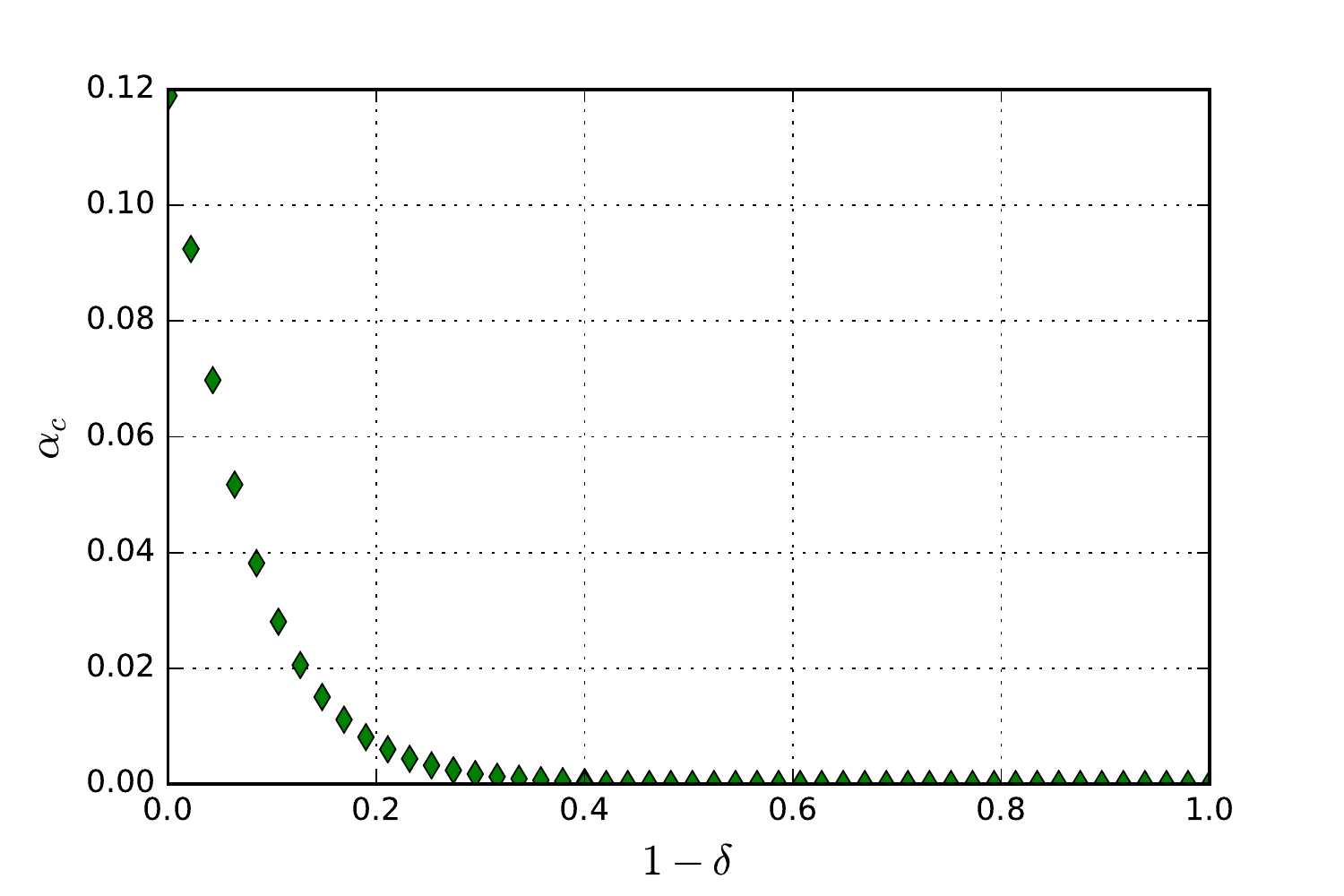}
\caption{Left panel: magnetization vs $\a$ for different values of $\d$; at $\a_c(\d)$ a first order phase transition occurs. The low-load pattern overlap $m(\a=0;\d)$ tends to $\frac{2}{\pi}$ as $\d\to 0$. Right panel: $\a_c(\d)$ plotted versus $1-\d$: $\a_c(1)=0.12..$, while rapidly $\a_c(\d)\to 0$ as $\d \to 0$.  }\label{fig2}
\end{figure}

For fixed $\d$, a first order phase transition occurs in the self-consistency condition (\ref{eq:x}) as $\a$ increases. The transition value $\a_c(d)$ is largest for $\d=1$ and decreases to zero quite rapidly as $\d\to0$, see Fig.~\ref{fig2}. For $\a<\a_c(\d)$ a non-zero solution of ($\ref{eq:x}$) exists, with $x$ (thus $m$) growing as $\a$ decreases. In particular, as $\a\to0$, $x=F_{\d,\a}(x)\to 1/\sqrt{2(1-\d^2)} = 1/\sqrt{2\Omega_\xi}$. In this low-load limit one then recovers for $m$ the expression  $(\ref{eq:ma0t0})$ as we show below.
 
We remark that since for any $0<\a<1$, $F_{\d,a}(x)\to0$ as $\d \to 0$, one also has $m\to0$ (with a first order phase transition, see Fig.~$\ref{fig2}$). For $\a=0$, on the other hand, we see from ($\ref{eq:m_low_T}$) that $m\to\sqrt{2/\pi}$ as $\d\to0$, which is consistent with the data shown in Fig.~$\ref{fig2}$. Thus the Hopfield model retrieves Gaussian patterns only for $\a=0$, but not at high load. 

We close this section by outlining the derivation of (\ref{eq:x}). Bearing in mind $\d=\sqrt{1-\Omega_\xi}$,  the equation (\ref{eq:m_low_T}) for $m$ becomes
\be\label{eq:mx}
m=\d \operatorname{erf} (\d  \,x) + (1-\d^2) \frac{2}{\sqrt{\pi}}x e^{-\d^2x^2}\,,
\ee
while for $C$ one gets
\be
C=\frac{2}{\sqrt{\pi}}\frac{x}{\Omega_\tau m}  \exp(-\d^2x^2)\,.
\ee
Thus
\begin{eqnarray}
V&=& \Omega_\tau^2 (1-\d^2) m^2 + \frac{\a}{1-\a} [ \Omega_\tau^{-1} -2x/(\sqrt{\pi}\Omega_\tau m) \exp(-\d^2 x^2)   ]^{-2}\nonumber \\
&=& \Omega_\tau^2 m^2 \{ 1-\d^2 +\frac{\a}{1-\a} [m- (2/\sqrt \pi) x\exp(-\d^2 x^2)     ]^{-2}    \} \nonumber \\
&=&  \Omega_\tau^2 m^2 \{ 1-\d^2 +\frac{\a}{1-\a} [\d\operatorname{erf} (\d  x) - (2/\sqrt \pi) \d^2 x\exp(-\d^2 x^2)    ]^{-2}    \}\,,
\end{eqnarray}
Now we set
\be 
F_{\d,\a}(x)=\{ 2(1-\d^2) + 2(1-\a) [\d \operatorname{erf} (\d x) - \frac{2}{\sqrt{\pi}}x\d^2 e^{-\d^2x^2}]^{-2} \}^{-1/2}\,.
\ee
and we readily get (\ref{eq:x}).

\subsection{Soft models}\label{ss:soft}
\begin{figure}
\includegraphics[scale=0.6]{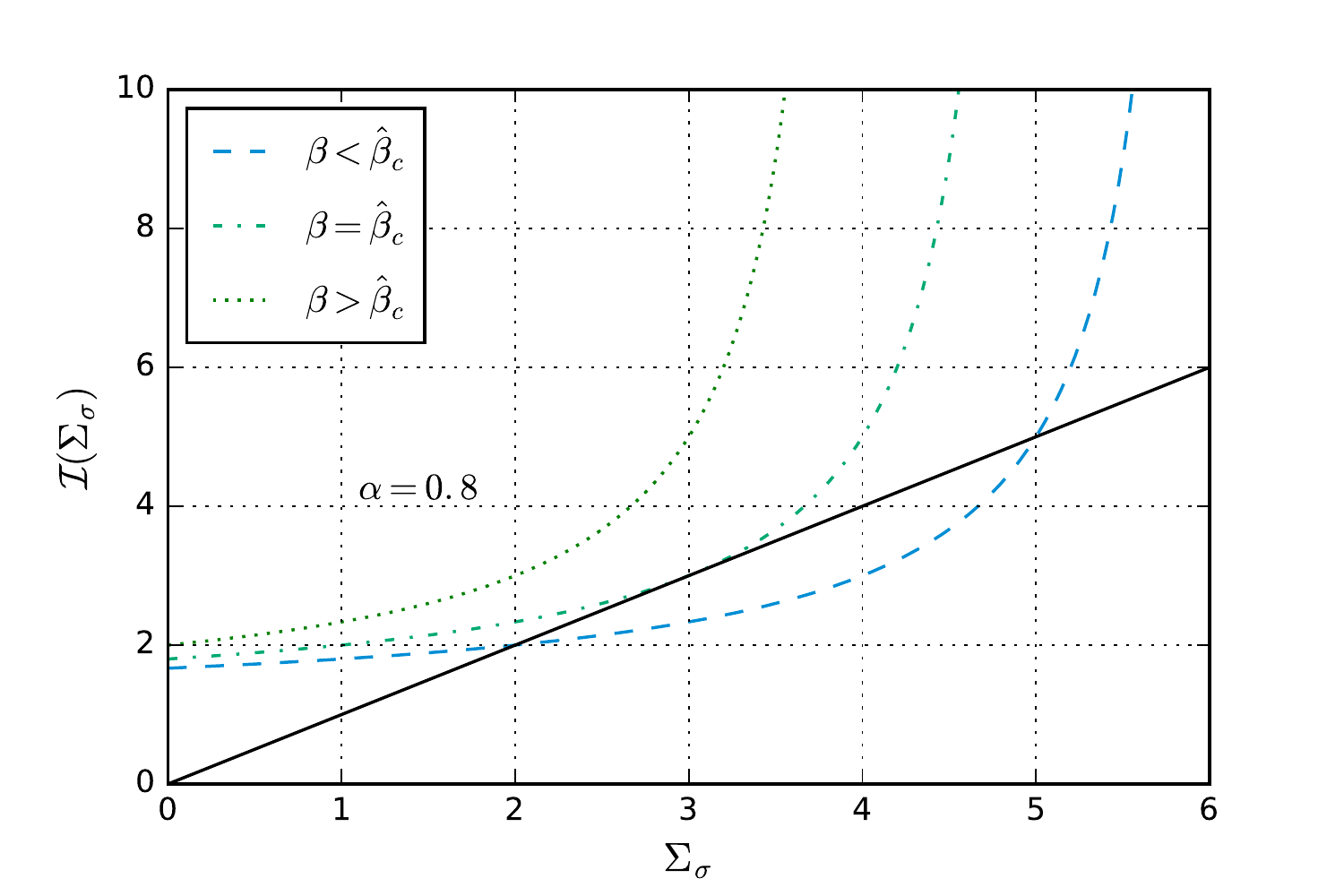}
\caption{$\mathcal{I}(x)$ vs $x$ for different value of $\b$. At $\hat{\b}_c$, $\mathcal{I}(x)$ is tangent to $x$.}\label{fig:I}
\end{figure}

Models with both Gaussian spins are typically ill-defined at low temperature, due to the occurrence of negative eigenvalues in the interaction matrix. In the fully Gaussian model ($\Omega_\s=\Omega_\tau=1$) the line where the partition function diverges coincides exactly with the paramagnetic/spin glass transition. In this case, the distributions $P(\s|z,\xi)$ and $P(\tau|\eta,\xi)$ of equations ($\ref{eq:effdistr1}$ - $\ref{eq:effdistr2}$) are respectively proportional to
\begin{eqnarray}
&&  e^{\b(1-\a)\Omega_\tau  m \xi\s+\sqrt{ \b\a r} z\s-\frac 1 2 (1-\b \a(R-r))\s^2}\\
&&  e^{ \b\a\Omega_\s  n\xi \tau+\sqrt{\b (1-\a) q}\tau \eta -\frac 1 2 (1-\b (1-\a) (Q-q)) \tau^2 }.
\end{eqnarray}
Both these distributions are therefore Gaussian with variances $\Sigma_\s$, $\Sigma_\tau$, defined by $\Sigma^{-1}_\s=1-\b\a(R-r)$ and $\Sigma^{-1}_\tau=1-\b (1-\a)(Q-q)$. The equations for $R$ and $Q$ read
\begin{eqnarray}
Q     &=& \meanv{ \meanv{\s^2}_{\s|z,\xi}}_{z,\xi}= q + \Sigma_\s  \\
R     &=&   \meanv{ \meanv{\tau^2}_{\tau |\eta}}_{\eta,\xi} = r + \Sigma_\tau \,.
\end{eqnarray}
Thus
\begin{eqnarray}
\Sigma_\s &=& \frac{1}{1-\b\a\Sigma_\tau}\nonumber \\
\Sigma_\tau &=& \frac{1}{1-\b(1-\a)\Sigma_\s}\,.
\end{eqnarray}
and one has to study the equation
$\mathcal{I}(\Sigma_\s)=\Sigma_\s$
where
\be \label{vareq}
\mathcal{I}(\Sigma_\s)= \frac{1-\b(1-\a) \Sigma_\s}{1-\b\a-\b(1-\a)\Sigma_\s}\,.
\ee
The function $\mathcal{I}(x)$ is a hyperbola diverging at $x=(1-\b\a)/\b(1-\a)$, see Fig.~\ref{fig:I}. 
It is positive only for $x$ below this value, so this is the range we need to consider as 
$\Sigma_\s
>0$. For small $\b$ one has a solution near $\Sigma_\s=1$ which increases with $\b$. At some $\hat{\b}_c$, $\mathcal{I}(x)$ becomes tangent to $x$ and for still larger $\beta$ there are no intersections.
After some calculations using ($\ref{vareq}$) one finds for the threshold $\hat{\b}_c$
\be
1=\hat{\b}_c^2\a(1-\a) \Sigma_\s^{2} \left(\frac{1}{1-\b(1-\a)\Sigma_\s}\right)^2=\hat{\b}_c^2\a(1-\a) \Sigma_\s^{2} \Sigma_\tau^{2},
\ee
which exactly coincides with the paramagnetic / spin glass transition temperature $(\ref{eq:tc})$ as anticipated.

We note that we can compute the divergence of the partition function of the model also directly, by diagonalising the interaction matrix (\ie the weight matrix):
\begin{eqnarray}
Z_N(\b,\a;\xi)&=&\E_{\s,\tau} \exp\left( \sqrt{\frac \b N} \sum_{i=1}^{N_1}\sum_{\mu=1}^{N_2} \xi^{\mu}_i \s_i\tau_\mu\right)\nonumber\\
&=&\E_{\s} \exp\left( \frac {\b} {2N} \sum_{i,j=1}^{N_1}\sum_{\mu=1}^{N_2} \xi^{\mu}_i\xi^{\mu}_j \s_i\s_j\right)=\E_{\s} \exp\left( \frac {\b\a} {2} \sum_{i,j=1}^{N_1}M_{ij}\s_i\s_j\right)
\end{eqnarray}
Here $\boldsymbol{M}= \frac 1 {N_2} \boldsymbol{\xi\xi}^\mathrm{T}$ is a Wishart matrix so its  empirical eigenvalue spectrum converges to the Marchenko-Pastur distribution for large $N_1$, which is nonzero only between $\left(1\pm\sqrt{{(1-\a)}/{\a}}\right)^2$. Using a suitable orthogonal transformation on the spin variables we can diagonalise $\boldsymbol{M}$, so that
$$
Z_N(\b,\a;\xi)=\E_{\s} \exp\left( \frac {\b\a} {2} \sum_{i}^{N_1}\l_i\s_i^2\right)\,.
$$
This is well-defined as long as $\max_i\left[\b(1-\a)\l_i\right]<1$. Using the largest eigenvalue from Marchenko-Pastur, 
$\max_i \l_i = (1+\sqrt{(1-\a)/\a})^2$ for large $N$, we get for the critical temperature
$$
T_c(\a)=(\sqrt{\a}+\sqrt{1-\a})^2\,.
$$
It can be checked that the spin glass transition line numerically computed in Section \ref{sect:SG-trans} coincides with $T=T_c(\a)$. In the general case $0<\Omega_\s,\Omega_\t<1$ we simply remark that (recall that the $g$ are $\mathcal N(0,1)$ and $\e=\pm1$)
\bea
\sum_{i=1}^{N_1}\sum_{\mu=1}^{N_2} \xi^{\mu}_i \s_i\tau_\mu&=&\sqrt{(1-\Omega_\s)(1-\Omega_\t)}\sum_{i=1}^{N_1}\sum_{\mu=1}^{N_2} \xi^{\mu}_i \epsilon_i\epsilon_\mu\\
&+&\left(\sqrt{\Omega_\t(1-\Omega_\s)}+\sqrt{\Omega_\s(1-\Omega_\t)}\right)\sum_{i=1}^{N_1}\sum_{\mu=1}^{N_2} \xi^{\mu}_i g_i\epsilon_\mu\\
&+&\sqrt{\Omega_\s\Omega_\t}\sum_{i=1}^{N_1}\sum_{\mu=1}^{N_2} \xi^{\mu}_i g_ig_\mu\,.
\eea
Of course the first two addenda have well-defined thermodynamical properties for all $T$, so we just need to rescale $T_c$ as
\be
T_c(\a)=\Omega_\s\Omega_\tau(\sqrt{\a}+\sqrt{1-\a})^2\,.
\ee
This generalises what happens at low load, where a divergence in $m$ appears at $T_c=\Omega_{\s}\Omega_{\tau}$ (see Fig.~\ref{fig:Gaussa0}). Note that such a critical temperature is lower than the one for the paramagnet/ spin glass transition. 

\subsection{Spherical Constraints}\label{ss:spherical-constr}

As before we can remove the singularity in the partition function by adding the spherical constraint $\d(N-\sum_{i=1}^N \s_i^2)$ to the $\s$-prior. The equations $(\ref{eq:gen})$ remain valid with the replacement
$$
\g_\s^{-1}=\Omega_\s^{-1}-\b\a(R-r)+\omega\,,   %ALERT
$$ 
with $\omega\geq0$ (or directly $\g_\s$, see also Section 3) satisfying
\be\label{eq:gamma2}
Q= q + \g_\s +\g_\s^2\phi_\s^2\left[1 -
\tsqav(\b(1-\a)\gamma_\s \phi_\s \Omega_\tau \d m,
\sqrt{v})\right]=1.
\ee
For binary $\sigma$, \ie $\Omega_\s\to 0$, one has $\g_\s\phi_\s\to 1$ and the constraint ($\ref{eq:gamma2}$) is automatically satisfied. For Gaussian $\sigma$ ($\Omega_\s=1$), on the other hand, $\phi_\s=0$ and hence $\g_\s=1-q$.

\begin{figure}
\includegraphics[scale=0.6]{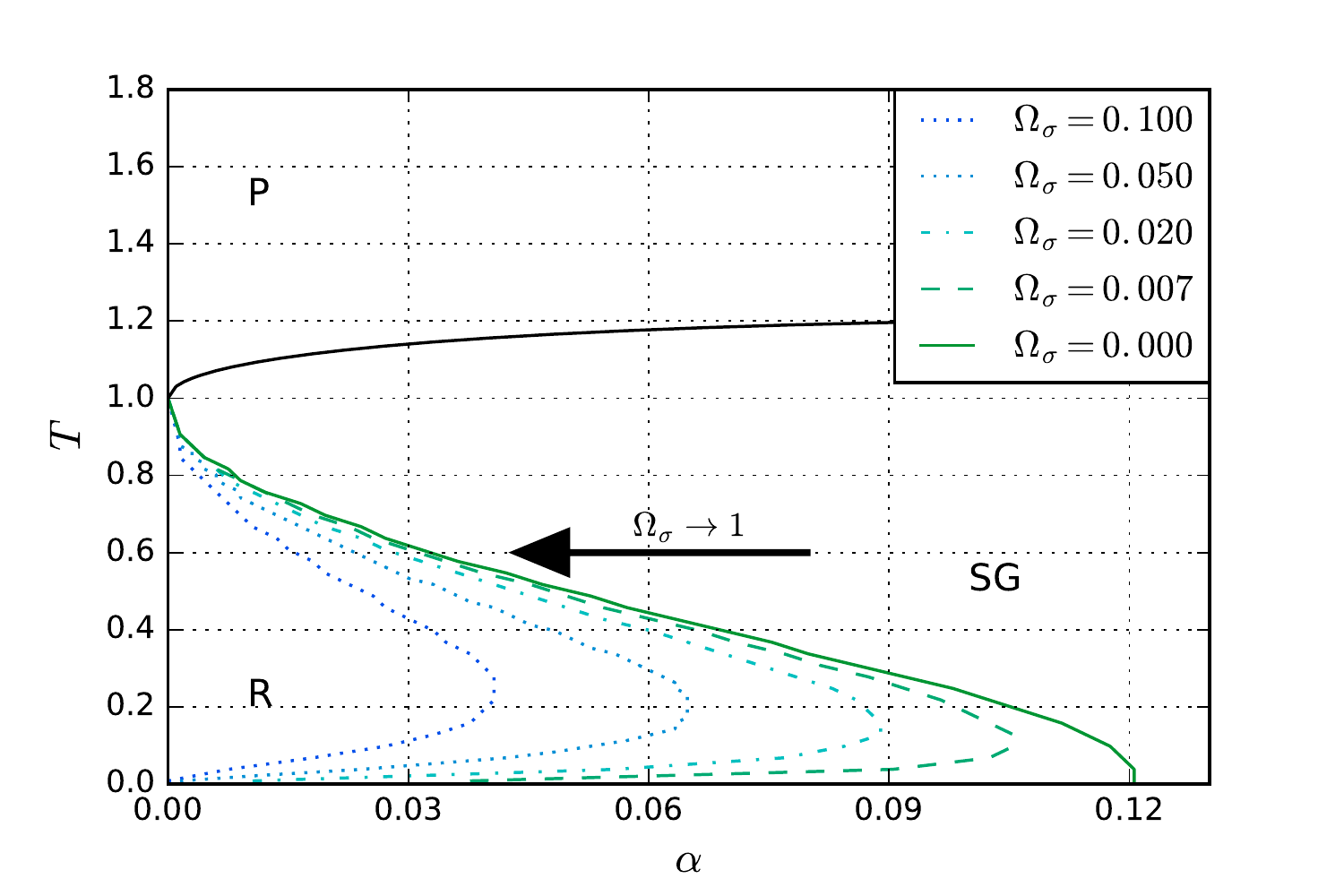}
\caption{Retrieval regions of the soft model with a spherical constraint on the $\s$-layer for different $\Omega_\s$ and fixed $\Omega_\tau=\delta=1$. }\label{fig:softcr}
\end{figure}

Starting from the low-load solution $\a=0$ and increasing $\alpha$, it is possible to find numerically the solution of the equations ($\ref{eq:gen}$) and the constraint ($\ref{eq:gamma2}$). The results, presented in Fig.~$\ref{fig:softcr}$, indicate that the retrieval region is robust also in the high-load regime, disappearing as $\Omega_\s\to1$. The retrieval transition line
exhibits re-entrant behaviour as in the standard Hopfield model, which might point to underlying RSB effects \cite{reentrant}.

In principle one can ask further what happens in a model where {\em both} layers have a spherical constraint. In this case we simply need to put an additional Gaussian factor $e^{-\omega_\tau \tau^2/2 }$ into the effective $\tau$-spin distribution, where the additional  Lagrange multiplier $\omega_\tau$  can be found by fixing the radius $R=1$.  As a consequence, the paramagnetic to spin glass transition line (\ref{eq:tc}) becomes
\be\label{eq:sfsfcr}
\b^2\a(1-\a)Q^2R^2=\b^2\a(1-\a)=1.
\ee
This is valid for the bipartite SK model ($\Omega_\s = \Omega_\t=0$) but also for generic $\Omega_\s$ and $\Omega_\tau$. 
As $T_c = \sqrt{\a(1-\a)}\to 0$ for $\a\to 0$ and retrieval is expected only {\em below} the paramagnetic to spin glass transition, this indicates that the double spherical constraint removes the possibility of a retrieval phase, even for low load. What is happening is that the high-field response $\Omega_\tau$  is weakened and becomes $\g_\tau^0=\Omega_\tau/(1+\Omega_\tau\omega_\tau)$. Moreover, equations $(\ref{eq:gen})$  still apply if we replace $\Omega_\tau$ by $\g_\tau^0$ and set $\g_\tau^{-1}=\Omega_\tau^{-1}-\b(1-\a)(1-q)+\omega_\tau$. In the paramagnetic regime $\g_\s$ and $\g_\tau$ satisfy
\begin{eqnarray}
Q= \g_\s +\g_\s^2\phi_\s^2=1 \ \ &\to& \ \ \g_\s=\Omega_\s\nonumber\\
R=  \g_\tau +\g_\tau^2\phi_\tau^2=1\ \ &\to& \ \  \g_\tau=\Omega_\tau,
\end{eqnarray}
while $q=0$, giving for the response $\g^0_\tau=1/(\g_\tau^{-1}+\b(1-\a))=(\Omega_\tau^{-1}+\b(1-\a))^{-1}$. This is not sufficient for retrieval, not even at low load ($\a=0$) where $\b\g^0_\tau=\b\Omega_\tau/(1+\b\Omega_\tau)<1$ and the critical temperature is $T=0$ ($\b\to\infty$). Intuitively, because of the spherical cut-off the tail of the hidden units is simply not sufficient to give, after marginalising out the visible units, an appropriate function $u$ (see Section \ref{ss:hopintro1})
to get spontaneous magnetisation in the low load ferromagnetic model.

\section{Conclusions and outlooks}

In this paper we have investigated the phase diagram of Restricted Boltzmann Machines with different units and weights distributions, ranging from centred (real) Gaussian to Boolean variables. We highlighted the retrieval capabilities of these networks, using their duality with generalised Hopfield models.

Our analysis is mainly based on the study of the self-consistency relations for the order parameters and offers a nearly complete description of the properties of these systems. For this rather large class of models we have drawn the phase diagram, which is made up of three phases, namely paramagnetic, spin glass and retrieval, and studied the phase transitions between them.

We stress that, while in associative neural networks patterns are often restricted to the binary case, there is at present much research activity in the area of Boltzmann machines with real weights. Our analysis shows that retrieval is possible at high load for any pattern distribution interpolating between Boolean and Gaussian statistics. In this Gaussian case high load retrieval fails, but is recovered even here at low load. 

A complete analysis of the paramagnetic-spin glass transition and the spin glass-retrieval transition is very useful for the study of modern deep neural networks, where the crucial learning phase is often initiated with a step of unsupervised learning through Restricted Boltzmann Machines \cite{DL-book, deeplearning}. A first attempt to link the properties of the phase diagram to the challenges of training a Restricted Boltzmann Machines from data and extracting statistically relevant features can be found in \cite{lettera}.

\vspace{1cm}

{\bf Acknowledgements}

A.B.\ acknowledges partial financial support by National Group of Mathematical Physics GNFM-INDAM G.G.\ is supported by the NCCR SwissMAP. D.T.\ is supported by Scuola Normale Superiore and National Group of Mathematical Physics GNFM-INDAM.

%%%%%%%%%%%%%%%%%%%%%%%%%%%%%%%%%%%%%%%%%%%%%%%%%%%%%%%%%%%%%

\newpage

\appendix

\section{Derivation of equations (\ref{eqs:0})-(\ref{eqs:5})}\label{App-replica}
Consider a bipartite system with $N_1$ $\s$-spins and $N_2$ $\tau$-spins, $N=N_1+N_2$, $\a=N_2/N$ and partition function
\be
Z_N(\b,\a;\xi)=\E_{\s,\tau} \exp\left( \sqrt{\frac \b N} \sum_{i=1}^{N_1}\sum_{\mu=1}^{N_2} \xi^{\mu}_i \s_i\tau_\mu\right),
\ee
with the expectation being over generic spin distributions $P_\s(\s)$ and $P_\tau(\tau)$.  We assume there are $\ell_1=O(1)$ condensed patterns associated with the first $\ell_1$ $\s$-variables and similarly $\ell_2$ condensed patterns associated with the first $\ell_2$ $\tau$-variables, and two families of overlaps
\be
 m^\mu(\s)=\frac 1 {N_1} \sum_{i=\ell_1}^{N_1}\xi^\mu_i \s_i\,, \quad n^i(\tau)=\frac 1 {N_2}\sum_{\mu=\ell_2}^{N_2}\xi^\mu_i \tau_\mu \,,
\ee
and
\be
q_{\a\b}=\frac 1 {N_1} \sum_{i=\ell_1}^{N_1}\s^\a_i\s^\b_i \quad r_{\a\b}=\frac{1}{N_2} \sum_{\mu=\ell_2}^{N_2}\tau^\a_\mu\tau^\b_\mu.
\ee
Then
\begin{eqnarray}
Z_N(\b,\a;\xi)&=& \E_{\s,\tau} \exp\left( \sqrt{\frac \b N} \sum_{i=1}^{\ell_1}\sum_{\mu=\ell_2}^{N_2}\xi^{\mu}_i \s_i\tau_\mu
+\sqrt{\frac \b N} \sum_{\mu=1}^{\ell_2}\sum_{i=\ell_1}^{N_1} \xi^{\mu}_i \s_i\tau_\mu\right)\nonumber \\
&\times&\exp\left( \sqrt{\frac \b N} \sum_{i=\ell_1}^{N_1}\sum_{\mu=\ell_2}^{N_2}\xi^{\mu}_i \s_i\tau_\mu+ \sqrt{\frac \b N} \sum_{i=1}^{\ell_1}\sum_{\mu=1}^{\ell_2} \xi^{\mu}_i \s_i\tau_\mu\right)\label{eqsim:1} \\
&\sim& \E_{\s,\tau} \exp\left( N_2\sqrt{\frac \b N} \sum_{i=1}^{\ell_1}n^i(\tau) \s_i
+N_1\sqrt{\frac \b N} \sum_{\mu=1}^{\ell_2}m^\mu(\s)\tau_\mu+\sqrt{\frac \b N} \sum_{i=\ell_1}^{N_1}\sum_{\mu=\ell_2}^{N_2}\xi^{\mu}_i \s_i\tau_\mu\right)\nonumber
\end{eqnarray}
where we have neglected the last, non-extensive, term of (\ref{eqsim:1}). Constraining the values of the overlaps we get 
\begin{eqnarray}
Z_N&=& \int \{dm^\mu, d\hat{m}^\mu,dn^i, d\hat{n}^i\}  \exp\left(-iN \left(\sum_{i=1}^{\ell_1} n^i \hat{n}^i +\sum_{\mu=1}^{\ell_2}m^\mu \hat{m}^\mu\right)\right)\nonumber\\
&\times&\E_{\s,\tau}\exp\left(N_2\sqrt{\frac \b N} \sum_{i=1}^{\ell_1}n^i \s_i+N_1\sqrt{\frac \b N} \sum_{\mu<\ell_2}m^\mu\tau_\mu\right)
\label{eqsim:3}
\label{eqsim:2}
\\
&\times& \E_{\s,\tau} \exp\left(\frac i {\a} \sum_{i=1}^{\ell_1} \hat{n}^i \sum_{\mu=\ell_2}^{N_2}\xi^\mu_i \tau_\mu + \frac i {1-\a} \sum_{\mu=1}^{\ell_2}\hat{m}^\mu \sum_{i=\ell_1}^{N_1} \xi^\mu_i \s_i
+\sqrt{\frac \b N} \sum_{i=\ell_1}^{N_1}\sum_{\mu=\ell_2}^{N_2}\xi^{\mu}_i \s_i\tau_\mu 
\right) \,.\nonumber
\end{eqnarray}

We recall the definition of $\Omega_{\s,\tau}$ and $u_{\s,\tau}$ from the Introduction: $u_{\s,\tau}$ is the cumulant generating function of $P_{\s,\tau}$, to wit $u_{\s,\tau}(h)=\ln\E_{P_{\s,\tau}}[e^{hx}]$ and
\be
\lim_{N\to\infty }\frac 1 N u_{\s,\tau}(\sqrt{N}x)=\frac {\Omega_{\s,\tau} x^2}{2}\,.
\ee
Then the terms in the second line of $(\ref{eqsim:2})$ become
\be
\E_{\s,\tau}\exp\left(N_2\sqrt{\frac \b N} \sum_{i=1}^{\ell_1}n^i \s_i+N_1\sqrt{\frac \b N} \sum_{\mu<\ell_2}m^\mu\tau_\mu\right)=\exp\left(\frac {\b N}{2} \left(\a^2 \Omega_\s \sum_{i=1}^{\ell_1} n_i^2+(1-\a)^2 \Omega_\tau \sum_{\mu=1}^{\ell_2}m_\mu^2 \right)    \right),
\ee
while, after introducing replicas and averaging over the disorder, the last term in (\ref{eqsim:3}) gives (with $u_\xi$ the cumulant generating function associated with the patterns)
\bea
\E_\xi \exp \left(\sqrt{\frac \b N} \sum_{\a=1}^n \sum_{i=\ell_1}^{N_1}\sum_{\mu=\ell_2}^{N_2}\xi^{\mu}_i \s^\a_i\tau^\a_\mu \right)&=& \exp \left(  \sum_{i=\ell_1}^{N_1}\sum_{\mu=\ell_2}^{N_2}u_\xi\left(\sqrt{\frac \b N} \sum_{\a=1}^n \s^\a_i\tau^\a_\mu\right)  \right)\nn\\
&\sim&  \exp \left(  \sum_{i=\ell_1}^{N_1}\sum_{\mu=\ell_2}^{N_2}\frac \b {2N} \sum_{\a,\b=1}^n \s^\a_i\s^\b_i\tau^\a_\mu\tau^\b_\mu \right)\,.\nonumber
\eea
(Here we have used that the patterns have unit variance, hence $u_\xi(x) = x^2+ \ldots$, and neglected corrections in $1/N$.)
This term becomes $\exp\left(\frac{\b N}{2} \a (1-\a) \sum_{\a\b} q_{\a\b} r_{\a\b} \right)$ once it is expressed in terms of the order parameters $q$ and $r$, bearing in mind that the {\em missing} spins $\s_1,\ldots,\s_{\ell_1}$ and $\tau_1,\ldots,\tau_{\ell_2}$ constitute a vanishing fraction of the total number. Now averaging over spin variables we get the other two terms in the last line of (\ref{eqsim:2}), where we also include the contributions from constraining the $q$ and $r$ order parameters: 
\begin{eqnarray}
&& \E_\s  \exp\left(  \frac i {1-\a}\sum_{\a=1}^n \sum_{\mu=1}^{\ell_2}\hat{m}^\mu_\a \sum_{i=\ell_1}^{N_1} \xi^\mu_i \s^\a_i   +\frac i {1-\a} \sum_{\a,\b=1}^n \hat{q}_{\a\b} \sum_{i=\ell_1}^{N_1} \s^\a_i\s^\b_i\right)\nonumber\\
&=& \exp\left( N(1-\a) \meanv{ \ln \E_\s e^{  \frac i {1-\a}(\sum_{\a=1}^n \sum_{\mu=1}^{\ell_2}\hat{m}^\mu_\a \xi^\mu \s^\a   + \sum_{\a,\b=1}^n \hat{q}_{\a\b} \s^\a\s^\b)}       }_\xi \right) \nonumber
\end{eqnarray}
and
\begin{eqnarray}
&& \E_\tau  \exp\left(  \frac i {(1-\a)}\sum_{\a=1}^n \sum_{i=1}^{\ell_1} \hat{n}^i_\a \sum_{\mu=\ell_2}^{N_2}\xi^\mu_i \tau^\a_\mu   +\frac i {(1-\a)} \sum_{\a,\b=1}^n \hat{r}_{\a\b} \sum_{\mu=\ell_2}^{N_2}\tau^\a_\mu\tau^\b_\mu\right)\nonumber\\
&=& \exp\left(  \a N \meanv{ \ln \E_\tau e^{  \frac i {\a}(\sum_{\a=1}^n \sum_{i=1}^{\ell_1} \hat{n}^i_\a \xi^i \tau^\a   + \sum_{\a,\b=1}^n \hat{r}_{\a\b} \tau^\a\tau^\b)}       }_\xi \right) \nonumber.
\end{eqnarray}
Collecting all the terms we get an expression for $\E[Z_N^n]$ which depends on the parameters $m^\mu_\a$, $n^i_\a$, $q_{\a\b}$ and $r_{\a\b}$:
\be
\E[Z_N^n] = \int \{dm^\a_\mu, d\hat{m}^\a_\mu\}\{dq_{\a\b},d\hat{q}_{\a\b}\} e^{N f(\{m_\a^\mu\},\{ n^i_\a\}, \{q_{\a\b}\},\{r_{\a\b}\} )},
\ee
with
\begin{eqnarray}\label{eqsim:f}
f(\{m^\mu_\a\}, \{n^i_\a\}, \{q_{\a,\b}\},\{r_{\a\b}\})&=& -\frac \b 2 \Omega_\tau (1-\a)^2 \sum_{\mu=1}^{\ell_2}{m^\mu_\a}^2-\frac \b 2 \Omega_\s \a^2 \sum_{i=1}^{\ell_1} {n^i_\a}^2 - \frac \b 2 \a(1-\a) \sum_{\a,\b=1} q_{\a\b} r_{\a\b}\nonumber \\
&+& (1-\a) \meanv{ \ln \E_\s e^{  \b(1-\a) \Omega_\tau \sum_{\a=1}^n (m\cdot\xi) \s^\a   + \frac{\b\a}{2} \sum_{\a,\b=1}^n r_{\a\b} \s^\a\s^\b     }       }_\xi \nonumber \\
&+& \a \meanv{ \ln \E_\tau e^{ \b\a \Omega_\s  \sum_{\a=1}^n (n\cdot\xi)\tau^\a   +\frac{\b(1-\a)}{2} \sum_{\a,\b=1}^n q_{\a\b} \tau^\a\tau^\b    }       }_\xi
\end{eqnarray}
By a saddle point calculation we obtain immediately
\be
i \hat{m}^\mu_\a=\b(1-\a)^2\Omega_\tau m^\mu_\a \quad i \hat{n}^i_\a=\b \a^2\Omega_\s n^i_\a \quad i\hat{q}_{\a\b}=\frac \b 2 \a (1-\a) r_{\a\b} \quad i\hat{r}_{\a\b}=\frac \b 2 \a (1-\a) q_{\a\b}
\ee
and in the RS ansatz, assuming that
\be
m^\mu_\a = m^\mu\quad n^i_\a = n^i \quad q_{ab} = Q \delta_{\a\b} + q (1-\delta_{\a,\b})\quad r_{ab} = R \delta_{\a\b} + r (1-\delta_{\a,\b}),
\ee
taking the limit $n\to 0$ and extremizing (\ref{eqsim:f}) we get the  saddle point equations in ($\ref{eqs:0}-\ref{eqs:5}$)

\section{Gaussian bipartite and spherical Hopfield model}\label{app:sf}
The bipartite system with Gaussian priors on both layers ($\Omega_\s=\Omega_\tau=1$) can be related to a spherical Hopfield model \cite{ST,IPC, baik-lee} via Legendre duality as in \cite{GT}. In fact, integrating over the radius $r\sqrt{N}$ we have
\begin{eqnarray}
Z_{g}(\b)&=&\int dr \frac{e^{-N{r^2}/{2}}}{\sqrt{2\pi}^N}\int d\Sigma_{r\sqrt{N}}(\s) e^{-\b \mathcal{H}(\s)}=\int dr \frac{e^{-N{r^2}/{2}}}{\sqrt{2\pi}^N} Z_{s}^{r\sqrt{N}}(\b)\nonumber\\
&=& \int dr \frac{e^{-N{r^2}/{2}}}{\sqrt{2\pi}^N} r^{N-1}\int d\Sigma_{\sqrt{N}}(\s) e^{-\b r^2 \mathcal{H}(\s)}=\int dr \frac{e^{-N{r^2}/{2}}}{\sqrt{2\pi}^N} r^{N-1} Z_{s}^{\sqrt{N}}(\b r^2),
\end{eqnarray}
where $d\Sigma_r(\s)$ is the uniform measure over the sphere of radius $r$ and ($Z_{g}$, $Z_{s}$) are respectively the partition functions of the Gaussian and spherical models. Thus the two free energies, $f_g$ and $f_s$, are related by
\be
-\b f_{g}= \sup_r \left[ -\frac 1 2 r^2 -\frac 1 2 \ln(2\pi) +\ln(r) -\b f_s(\b r^2)\right]
\ee
and so the Gaussian free energy comes from the spherical free energy calculated at the optimal radius given by
\be
r^2=\frac{1}{1-2\b\partial_\b(-\b f_s(\b))|_{\b r^2}}\,.
\ee
Since $r^2=Q$, the self overlap of the $s$-spins (first layer), and using the expression for the spherical free energy from \cite{IPC,baik-lee} we have, in the high-temperature region,
\be
Q=\frac{1}{1-\b\a\frac{1}{1-\b(1-\a) Q}}=\frac{1}{1-\b\a R(Q)}\quad \hbox{and}\quad R(Q)=\frac{1}{1-\b(1-\a) Q}\,.
\ee
These are exactly equations ($\ref{eqQ}$)-($\ref{eqR}$) with $\Omega_\s=\Omega_\tau=1$. Moreover, again from \cite{IPC,baik-lee} the critical line for the spherical model is given by $(1-\b(1-\a))^2=\b^2\a(1-\a)$. Thus we obtain the critical line for the Gaussian model ($\ref{eq:tc}$) by replacing $\b\to\b Q$: 
\be
\frac{\b^2\a(1-\a) Q^2 }{(1-\b(1-\a) Q)^2}= 1=\b^2\a(1-\a) Q^2 R^2\,.
\ee

\end{document}